\documentclass[prl,preprintnumbers,amsmath,amssymb,preprint]{revtex4}

\usepackage{graphicx}
\usepackage{dcolumn}
\usepackage{bm}
\usepackage{times}
\usepackage{amsmath}
\usepackage{amsfonts}
\usepackage{amssymb}
\usepackage{color}

\newcommand{\m}[1]{\mathrm{#1}}

\definecolor{blue}{rgb}{0,0,1}

\begin{document}

\title{Dipolar and charged localized excitons in carbon nanotubes}

\author{Jan T. Gl\"uckert$^1$, Lyudmyla Adamska$^{2}$, Wolfgang Schinner$^1$, Matthias S. Hofmann$^1$, Stephen K. Doorn$^3$, Sergei Tretiak$^2$, and Alexander H\"ogele$^1$}

\affiliation{$^1$Fakult\"at f\"ur Physik, Munich Quantum Center,
and Center for NanoScience (CeNS),
Ludwig-Maximilians-Universit\"at M\"unchen,
Geschwister-Scholl-Platz 1, D-80539 M\"unchen, Germany}

\affiliation{$^2$Theory Division, Los Alamos National Laboratory,
Los Alamos, New Mexico 87545, U.S.A}

\affiliation{$^3$Center for Integrated Nanotechnologies, Materials
Physics and Applications Division, Los Alamos National Laboratory,
Los Alamos, New Mexico 87545, U.S.A}

\date{\today}

\begin{abstract}
We study both experimentally and theoretically the fundamental
interplay of exciton localization and polarization in
semiconducting single-walled carbon nanotubes. From Stark
spectroscopy of individual carbon nanotubes at cryogenic
temperatures we identify localized excitons as permanent electric
dipoles with dipole moments of up to $1~e$\AA. Moreover, we
demonstrate field-effect doping of localized excitons with an
additional charge which results in defect-localized trions. Our
findings provide not only fundamental insight into the microscopic
nature of localized excitons in carbon nanotubes, they also
signify their potential for sensing applications and may serve as
guidelines for molecular engineering of exciton-localizing quantum
dots in other atomically thin semiconductors including transition
metal dichalcogenides.
\end{abstract}

\maketitle

Optical transitions of semiconducting carbon nanotubes (CNTs) are
dominated by excitons \cite{Wang2005,Maultzsch2005} which exhibit
strong antibunching in the photoluminescence (PL) in their
localized limit \cite{Hogele2008}. Exciton localization can arise
at unintentional defects with shallow potentials
\cite{Georgi2010,Hofmann2016} or incidental proximal charges
\cite{Tayo2012} and ensure non-classical emission statistics up to
room-temperature  \cite{Ma2015nn} for excitons bound to deep traps
of oxygen side-wall dopants \cite{Ghosh2010,Ma2014a}. Along with
oxygen functionalization \cite{Ghosh2010,Ma2014a,Ma2015adv},
covalent side-wall chemistry with aryl and alkyl functionality
\cite{Piao2013,Kwon2016} provides a versatile molecular means to
engineer the photophysics of semiconducting CNTs. Introduced in a
moderate concentration, the decoration of CNT side-walls with
covalent defects results in substantial modifications such as
brightening of nanotube emission and increased quantum yields
\cite{Ghosh2010,Miyauchi2013,Piao2013}, axially pinned PL
\cite{Hartmann2015} and inhibited diffusion
\cite{Ma2014b,Ma2015prl,Hartmann2016}.

Defect-localized excitons in CNTs represent a viable resource for
applications in quantum sensing and quantum cryptography. For the
latter technology, CNTs may facilitate the development of robust
single-photon sources with room-temperature operation in the
telecom band by utilizing discrete optical transitions of
defect-localized excitons \cite{Ma2015nn}. Covalent chemistry is
readily available to fine-tune the exciton PL energy
\cite{Ghosh2010,Ma2014a,Ma2015adv,Piao2013,Kwon2016}, and recent
successful integration of CNTs into optical cavities
\cite{Jeantet2016,Hummer2016} has demonstrated Purcell enhancement
and directional coupling of single-photon emission as means to
increase the single-photon emission efficiency. Moreover, the
interplay of chemical modification and charge doping facilitates
photoemission from trions
\cite{Matsunaga2011,Park2012,Brozena2014}, which can be utilized
to interface photons with the CNT spin degree of freedom
\cite{Galland2008spin} via schemes of spin-tagged optical
transitions analogous to charged semiconductor quantum dots and
nitrogen vacancy (NV) centers in diamond \cite{Gao2015}. This
spin-photon interface in turn should enable all-optical sensing of
magnetic fields in analogy to magnetometry based on charged NV
color centers \cite{Maze2008,Balasubramanian2008}. A corresponding
nanoscale sensor for the measurement of the electric field
\cite{Vamivakas2011} with sensitivity down to the elementary
electron charge \cite{Houel2012,Hauck2014} could utilize the
electric dipole moment associated with localized excitons.

Our work identifies both integral elements - dipolar localized
excitons and voltage-controlled trions - for the development of
sensing devices based on carbon nanotubes. By embedding CNTs in a
field-effect (FET) device, we performed Stark spectroscopy of
localized nanotube excitons in a transverse electric field at
cryogenic temperatures. Our experiments demonstrate that exciton
localization is accompanied by static exciton polarization
irrespective of the details of the localizing potential. An
average localization-induced electric dipole moment of $\sim
0.3~e$\AA~found experimentally is in good quantitative agreement
with ab-initio model calculations for excitons bound by oxygen
defects on the side-wall of a $(6,5)$ nanotube. Moreover, we found
that defect potential traps can bind an additional charge to
promote PL from defect-localized trions
\cite{Mouri2013,Brozena2014}, with control over the charging state
provided by the gate voltage.

\begin{figure}[t]
\includegraphics[scale=1.09]{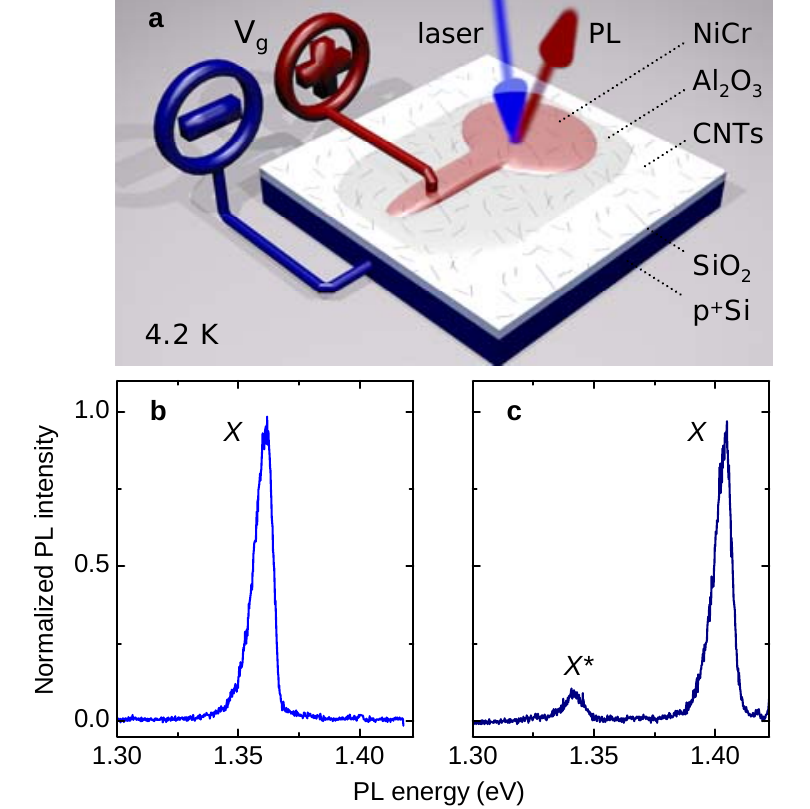}
\caption{(a) Schematic layout of the experiment: carbon nanotubes,
dispersed in between insulating oxide layers (SiO$_2$ and
Al$_2$O$_3$), were subjected to a transverse electric field by
applying a gate voltage $V_g$ to the semitransparent top gate
(NiCr) with respect to the back gate (p-doped Si). The response of
individual nanotubes to the electric field was studied with
confocal excitation (laser) and detection of the photoluminescence
(PL) at the temperature of liquid helium ($4.2$~K). (b) and (c),
Photoluminescence spectra of individual nanotubes with single-peak
($X$) and double-peak ($X$, $X^*$) emission spectra at $V_g=0$~V.}
\label{fig1}
\end{figure}

To subject nanotubes to a transverse electric field we fabricated
FET devices based on a metal-oxide-semiconductor sequence as
illustrated in Fig.~\ref{fig1}a. The FET devices were fabricated
starting with a $p^{+}-$doped silicon back gate terminated by an
insulating layer of $d_1=100$~nm thermal SiO$_2$ that was cleaned
with standard solvents and subsequently exposed to an oxygen
plasma before spin-coating micelle-encapsulated CoMoCat CNTs with
a spatial density below $\mu$m$^{-2}$.  The CNT layer was
subsequently covered by sputter deposition with an insulating
layer of Al$_2$O$_3$ of variable thickness $d_2$ (with $d_2=7$,
$17$, $39$ and $42$~nm in four different sample layouts), and a
semitransparent NiCr layer of $5$~nm thickness. A gate voltage
$V_g$ applied between the top and the ground electrode resulted in
a homogeneous transverse electric field $F$ through $F=V_g/d$,
with $d$ being the total thickness of the oxide layers. The
functionality of our FET devices with break-down voltages of $\pm
80$~V at low temperatures, corresponding to transverse electric
field strengths of up to $\pm 1~\m{V}/\m{nm}$, was confirmed with
capacitance-voltage spectroscopy \cite{SI}.

Individual CNTs embedded in a FET device were studied with
photoluminescence (PL) spectroscopy in a home-built confocal
microscope at the temperature of liquid helium of $4.2$~K. A
Ti:sapphire laser tuned in the range of $730 - 900$~nm was used to
excite the PL via phonon sidebands in continuous wave mode. The PL
of individual CNTs was dispersed with a monochromator and recorded
with a low-noise nitrogen-cooled silicon CCD. Our experiments
focused on $(6,4)$ and $(9,1)$ chiral nanotubes with emission in
the spectral range of $1.35 - 1.43$~eV \cite{Bachilo2003}.
Characteristic PL signatures of individual nanotubes in our device
are shown in Fig.~\ref{fig1}b and c. Most of the CNTs were found
to exhibit either a single-peak PL emission with an asymmetric
lineshape (labelled as $X$ in Fig.~\ref{fig1}b) characteristic of
disorder-localized excitons \cite{Galland2008,Vialla2014} or a
two-peak emission spectrum (denoted as $X$ and $X^*$ in
Fig.~\ref{fig1}c). In our experiments, we assign one-peak spectra
to excitons localized by environmental disorder, and two-peak
spectra to exciton PL from oxygen-dopant sites introduced on CNT
side-walls by sputter deposition of Al$_2$O$_3$ \cite{Ma2015adv}.

\begin{figure*}[t]
\includegraphics[scale=1.09]{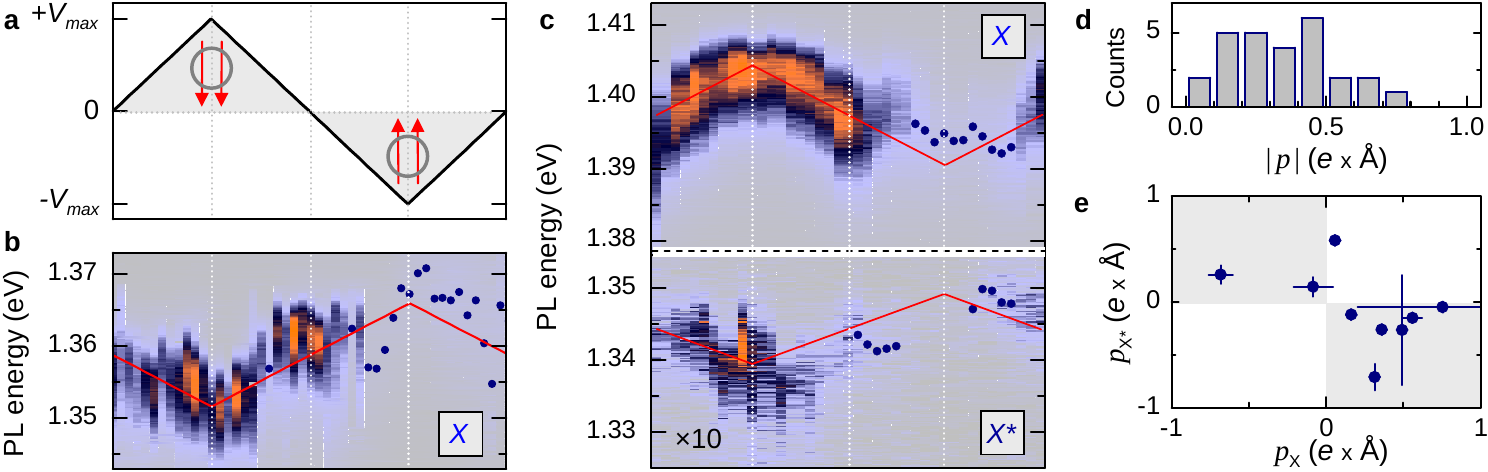}
\caption{(a) The transverse electric field, with orientation
indicated by the inset schematics, was varied via the gate voltage
between maximum negative and positive values according to the
zig-zag ramp. (b) and (c), False-color plots of the
photoluminescence from nanotubes in Fig.\ref{fig1} in response to
the field ramp. In the regions of low intensity the peak maxima
are shown as blue circles where peak fitting converged; the
intensity in the lower panel of (c) was magnified by a factor of
$10$. The red solid lines are linear fits to the dispersion with
slopes determined by the corresponding permanent dipole moments
$p$. (d) Histogram of the absolute values of dipole moments
extracted from linear fits as in (b) and (c) for nanotubes with
emission energy in the range $1.35-1.43$~eV. (e) For the majority
of nanotubes with double-peak emission, the signs of dipole
moments of the $X$ and $X^*$ states, $p_{X}$ and $p_{X^*}$, were
anticorrelated (data in the grey-shaded quadrants).} \label{fig2}
\end{figure*}

The evolution of the CNT spectra with a single-peak and a
double-peak spectrum as a function of the transverse electric
field are shown in Fig.~\ref{fig2}b and c, respectively. The
electric field strength and orientation was varied proportional to
the zig-zag voltage ramp shown in Fig.~\ref{fig2}a. The gate
voltage was changed in discrete steps between maximum positive and
negative values, with $V_{\mathrm{max}}$ ranging between $15$~V
and $30$~V depending on the device. After each voltage step a PL
spectrum was acquired for an incremental build-up of PL intensity
false-color plots as in Fig.~\ref{fig2}b for a single-peak
emission, and in Fig.~\ref{fig2}c for the $X$ and $X^*$ peaks. We
repeated this procedure on more than 50 individual CNTs. Roughly
one third of the tubes we have investigated showed irregular
responses such as non-monotonic energy jumps or irreversible
intensity fluctuations and were discarded from further analysis.
The more regular responses as in Fig.~\ref{fig2}b and c are
representative for CNT excitons localized by incidental and
oxygen-specific defect traps, respectively.

The vast majority of the nanotubes in our devices exhibited linear
energy dispersions in response to the transverse electric field
ramp, and both blue- and red-shifts were observed \cite{SI} for
different peaks (Fig.~\ref{fig2}b and c). The linear slope,
associated with the first-order Stark response of a permanent
dipole, is in striking contrast to the second-order Stark effect
expected for pristine CNTs. From a fitting procedure of CNT PL
with single- and double-peak emission spectra as a function of the
electric field strength according to $E(F)= E_0 - p F$ (red solid
lines in Fig.~\ref{fig2}b and c) we extracted the transverse
dipole moment $p$ of localized excitons with emission energy $E_0$
at $V_g=0$~V. For the CNT in Fig.~\ref{fig2}b, we obtained
$p_{X}=-0.38~e\mathrm{\AA}$, and for the two states $X$ and $X^*$
of Fig.~\ref{fig2}c we determined $p_{X}=0.36~e\mathrm{\AA}$ and
$p_{X^*}=-0.26~e\mathrm{\AA}$ from linear fits to the data.

The results of the fitting procedure for all other CNTs with
single- and double-peak emission are summarized in the histogram
of Fig.~\ref{fig2}d. It shows the distribution of the absolute
value of the transverse permanent dipole moments determined for
different CNTs and devices. The maximum value of the distribution
at $|p| \simeq 0.7~e\mathrm{\AA}$ corresponds to an electron-hole
separation of $\sim 10\%$ of the CNT diameter, a remarkably large
value for a permanent dipole moment that is absent in pristine
CNTs according to symmetry considerations. Another remarkable
trend in our data are the anti-correlated signs of the dipole
moments associated with $X$ and $X^*$ peaks (data points within
the grey-shaded quadrants in Fig.~\ref{fig2}e). Among the tubes
with two-peak spectra, the majority exhibited positive $p_{X}$ and
negative $p_{X^*}$ permanent dipole values (corresponding to data
points in the lower right quadrant).

Our experimental observations suggest an intimate interplay of
exciton localization and polarization which we confirmed by
atomistic calculations of a $(6,5)$ model nanotube in transverse
electric field \cite{SI}. In our calculations, the nanotube was
embedded in a homogeneous medium with permittivity
$\varepsilon_r=6.3$ to account for the effective dielectric
environment composed of Al$_2$O$_3$ ($\varepsilon_r=9.3$) and
Si$_2$O$_2$ ($\varepsilon_r=3.9$) layers at the top and bottom of
the tube and micellar encapsulation. First, we modelled the
response of a pristine tube and found a quadratic energy
dispersion of the bright luminescent state with transverse
polarizability $\alpha_{\perp} \simeq 7.7~\mathrm{\AA}^{2}$ in
accord with previous estimates both from tight-binding
\cite{Benedict1995,Li2003,Novikov2006} and first-principles
calculations \cite{Guo2004,Brothers2005,Kozinsky2006}.

\begin{figure}[t]
\includegraphics[scale=1.09]{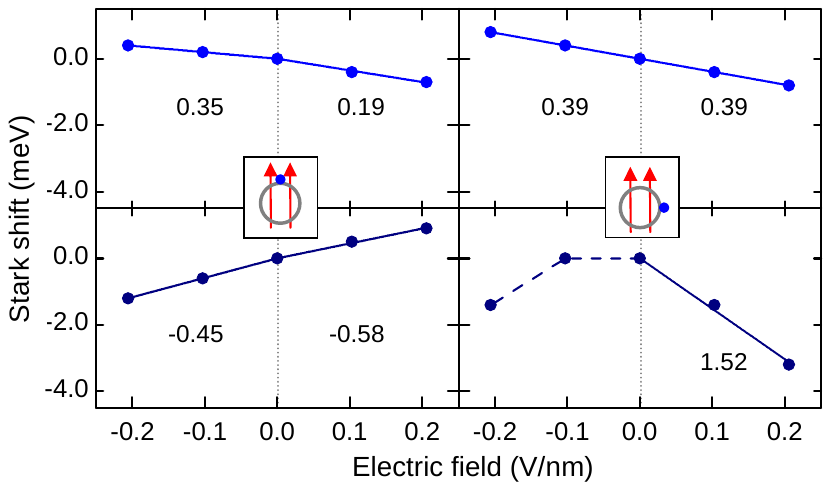}
\caption{Electric field dispersions of the two bright states (blue
and dark blue data points) associated with an oxygen-defect in
ether-d configuration and orientation as indicated in the inset
schematics. The numbers give the slopes in units of the permanent
dipole moment, $e\mathrm{\AA}$, extracted from linear fitting
(solid lines) performed separately for positive and negative
electric fields. The dashed line is a guide to the eye.}
\label{fig3}
\end{figure}

In stark contrast, for both bright peaks associated with an oxygen
side-wall defect in ether-d configuration \cite{Ma2014a}, our
calculations yield predominantly linear dispersions
(Fig.~\ref{fig3}) when subjected to a transverse electric field of
up to $0.2~$V/nm. The slopes and signs depend on the position of
the defect on the nanotube side-wall as indicated by inset
schematics in Fig.~\ref{fig3}. Our calculations predict red- and
blue-shifts for the $X$ and $X^*$ emission, respectively, with
corresponding maximum values for the permanent dipole moments of
$0.35~e\mathrm{\AA}$ and $-0.58~e\mathrm{\AA}$ for a defect placed
at the apex of the tube (left panel of Fig.~\ref{fig3}). This
defect geometry is expected to dominate our experiments with
side-wall dopants introduced preferentially from the top by oxide
sputtering, whereas localizing sites at the nanotube base caused
by proximal charges \cite{Tayo2012} at the SiO$_2$ surface should
be less frequent. Consistently, our experimental data of
Fig.~\ref{fig2}e reflects both the anti-correlated signs of the
two-peak dispersions predicted by theory, and the different
likelihood for defects to occur at the top and the bottom of the
tubes (in the latter case the respective slopes would remain
anti-correlated but interchange their signs). Experimental
observation of dispersions as in the right panel of
Fig.~\ref{fig3} should be rare because of the peripheral
configuration of the related defects in the top-down sputter
deposition process.

\begin{figure}[t!]
\includegraphics[scale=1.09]{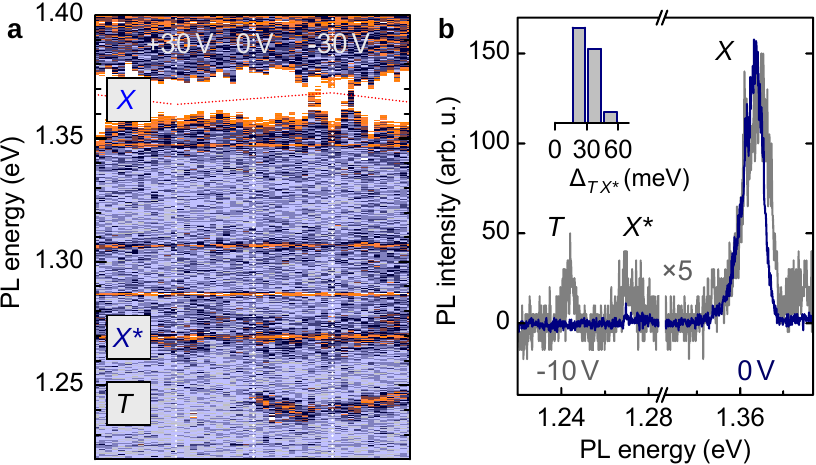}
\caption{(a) Logarithmic false-color plot of the photoluminescence
evolution in response to a field ramp as in Fig.~2a with
peak-voltage values of $\pm 30$~V. The narrow peaks are Raman
sidebands of the excitation laser at $1.484$~eV amplified by the
logarithmic color scale. In addition to the $X$ and $X^*$ peaks,
trion ($T$) satellite emission appears at negative gate voltages.
(b) Photoluminescence spectra at $V_g=0$~V (blue) and $-10$~V
(grey, magnified by a factor of 5). The distribution of trion
binding energies, as given by the energy splitting
$\Delta_{\mathrm{TX^*}}$ between the $T$ and $X^*$ peaks of all
nanotubes with trion satellites, is shown in the inset.}
\label{fig4}
\end{figure}

Both experiment and theory suggest that the radial symmetry of the
electron-hole charge distribution is imbalanced at the
exciton-localizing defect sites by field gradients associated with
defect traps, and both the strength and the orientation of the
respective dipole moment depend on the specifics of the localizing
defect. The defect potentials should also act as traps for
individual charges \cite{Brozena2014,Hartmann2015} and, in the
presence of photoexcited electron-hole pairs, give rise to
emission from energetically lower-lying trions
\cite{Matsunaga2011,Santos2011,Park2012,Mouri2013,Brozena2014}.
Indeed, we observed signatures of such red-shifted PL satellites
for some nanotubes within limited gate voltage ranges of our
devices. Fig.~\ref{fig4}a shows the PL response of a CNT to the
gate voltage ramp as in Fig.~\ref{fig2}a. The PL intensity is
represented on a logarithmic false-color scale to enhance the
visibility of the weak lowest-energy satellite which we assign to
defect-localized trion PL emission (denoted in Fig.~\ref{fig4}a as
$T$; the sharp horizontal features unaffected by the gate voltage
correspond to Raman scattered laser photons). For this specific
tube, the $T$ peak was observed around $1.24$~eV in addition to
$X$ and $X^*$ emission only at negative gate voltages. Other CNTs
exhibited similar features only for positive voltages \cite{SI}
indicating that the polarity of the defect excess charge trapped
out of the optically excited charge reservoir \cite{Santos2011}
depends on the defect potential details. Akin to previous
experiments
\cite{Matsunaga2011,Santos2011,Park2012,Mouri2013,Brozena2014},
the trion emission emerges at the expense of the main peak PL
intensity (compare the relative intensities of $T$ and $X$ peaks
at $0$~V and $-10$~V in Fig.~\ref{fig4}b).

Further confirmation for the assignment of the voltage-induced
satellite to trion emission comes from the inspection of the trion
binding energy. We extract the energy scale associated with the
binding of an excess charge to the lowest defect-localized state
by taking the energy splitting $\Delta_{\mathrm{TX^*}}$ between
the $T$ and $X^*$ emission peaks. This splitting, shown for all
CNTs with charging signatures in the inset histogram of
Fig.~\ref{fig4}b, varies between $20$ and $60$~meV for the $(6,4)$
and $(9,1)$ narrow-diameter tubes in the spectral region of our
experiment. This trion binding energy is not to be confused with
earlier experiments measuring the splitting between the trion peak
and the $E_{11}$ emission energy with excess contribution from
exchange interactions
\cite{Matsunaga2011,Santos2011,Park2012,Yoshida2016}. It should be
rather compared with the theoretical estimate of the bare trion
binding energy \cite{Ronnow2009}, or with the energy splitting
observed between the neutral and charged defect-localized emission
peaks in diazonium-functionalized CNTs \cite{Brozena2014}. Theory
predicts a trion binding energy of about $30$~meV for a $(6,5)$
nanotube in a dielectric medium with $\varepsilon_r =6$
\cite{Ronnow2009}. In aqueous suspension with $\varepsilon_r
\simeq 2$, the corresponding experimental value of $\sim 100$~meV
\cite{Brozena2014} was found in accord with the scaling of the
trion binding energy with the dielectric constant as
$\varepsilon_r^{-1.56}$ \cite{Ronnow2009}. Given the relatively
high effective dielectric constant of the CNT environment in our
FET devices and same diameters of $(6,5)$ and $(9,1)$ CNTs, we
find very good agreement between our lower values of
$\Delta_{\mathrm{TX^*}}$ and theory. Consistently, the larger
values in the distribution of Fig.~\ref{fig4}b are associated with
$(6,4)$ oxygen-doped nanotubes because of the inverse dependence
of the trion binding energy on the tube diameter
\cite{Ronnow2009}.

Our observation of defect-localized emission in combination with
voltage-controlled charging places CNTs alongside semiconductor
quantum dots \cite{Warburton2000} and NV centers \cite{Gruber1997}
with charge-tunable emission characteristics and spin-projective
optical transitions \cite{Gao2015}. An intriguing advantage of
CNTs for spin-based applications is expected to arise from
prolonged electron spin coherence time in an isotopically
engineered nuclear-spin free lattice
\cite{Balasubramanian2009iso}. Moreover, the absence of dangling
bonds in $sp^2$-hybridized CNTs could enable long spin coherence
times of electrons localized at engineered nanotube side-wall
defects with immediate environmental proximity - a key factor for
nanoscale-magnetometry \cite{Mamin2013,Staudacher2013} where
near-surface color centers in diamond currently encounter major
limitations due to unsaturated $sp^3$-bonds of the diamond crystal
surface \cite{Laraoui2012}. Finally, our results could inspire
efforts to create chemically engineered quantum dots for in-plane
confinement of excitons in emergent two-dimensional transition
metal dichalcogenide semiconductors \cite{Wang2012}.


We thank J.~P.~Kotthaus, S.~Rotkin, I.~Bondarev and V.~Perebeinos
for useful discussions, P.~Altpeter and R.~Rath for assistance in
the clean room, and P.~Stallhofer from Wacker~AG for providing the
wafer material. This work was performed in part at the Center for
Integrated Nanotechnologies, a U.S. Department of Energy, Office
of Science user facility. The research was funded by the European
Research Council (ERC) under the grant agreement No. 336749 and
the Deutsche Forschungsgemeinschaft (DFG) Cluster of Excellence
NIM (Nanosystems Initiative Munich) with financial support from
the Center for NanoScience (CeNS), LMUinnovativ, and the LANL LDRD
program.

\cleardoublepage

\setcounter{figure}{0} \setcounter{equation}{0}

\section{Supplementary online material}

{\bf Field-effect device characteristics}

In order to apply transverse electric fields to individual
nanotubes we fabricated metal-oxide-semiconductor (MOS) devices as
illustrated in Fig.~\ref{fig1}a. Highly $p^{+}-$doped silicon
substrate terminated by an insulating layer of $d_1=100$~nm
thermal silicon oxide (SiO$_2$) was used as the ground electrode.
The sample surface was cleaned with standard solvents and
subsequently exposed to an oxygen plasma. Commercial
CoMoCat-nanotubes (SouthWest NanoTechnologies) encapsulated in
sodium dodecylbenzenesulfonate (SDS) were dispersed out of an
aqueous suspension on SiO$_2$ substrates. The average CNT length
was $\sim 500$~nm, the spin coating parameters were adjusted to
yield a CNT density below $1~\mu$m$^{-2}$ which was confirmed by
AFM imaging like in Fig.~\ref{fig1}a. The CNT layer was
subsequently covered with a second insulating layer of aluminum
oxide (Al$_2$O$_3$) of variable thickness $d_2$, yielding a total
oxide thickness $d=d_1+d_2$. A semitransparent top electrode of
$3-5$~nm nickel chromium completed the MOS structure.

To determine the strength of the homogeneous electric field in our
MOS devices with $d_2=7$, $17$, $39$ and $42$~nm we performed
capacitance-voltage (CV) measurements at $4.2$~K. We used a
differential CV measurement technique by admixing a sinusoidal
modulation voltage with amplitude $\delta V=10$~mV and frequency
$f$ in the rage of $4-450$~Hz to the dc gate voltage. The
resulting ac capacitive current was demodulated with a lock-in
amplifier and scaled to the current of a reference capacitance. A
representative CV curve recorded for a device with $d_2=7$~nm is
shown in Fig.~\ref{fig1}b. At high negative voltages the MOS
device response was dominated by hole accumulation where we
obtained the maximum capacitance $C_i \simeq 1$~nF in accord with
the geometry of our device. For more positive values of $V_g$ the
CV curve showed a reduction of the capacitance to
$C_\textnormal{min}/C_i \simeq 0.93$ without recovery despite
further biasing and slow modulation (black CV traces in
Fig.~\ref{fig1}b), indicating that the limit of strong inversion
was suppressed in our device because of slow
generation-recombination rates of minority charge carriers at
$4.2$~K. This feature as well as the hysteresis at positive $V_g$
are characteristic for non-ideal $p$-type MOS-capacitors with
inhibited inversion \cite{Sze,Grove} and distinct charging and
discharging dynamics of charge traps at the $\m{Si-SiO}_2$
interface \cite{Goetzberger}. We note, however, that this
non-ideal capacitance change as a function of the gate voltage is
sufficiently small ($<10 \%$) to establish a linear relation
between the macroscopic electric field strength, $F$, and the gate
voltage, $V_g$, through $F=V_g/d$. With this relation we estimate
that field strengths of up to $\pm 1~\m{V}/\m{nm}$ were routinely
accessible with voltages of $\pm 80$~V applied to our devices at
low temperatures without break-down.

\begin{figure}[t]
\includegraphics[scale=1.10]{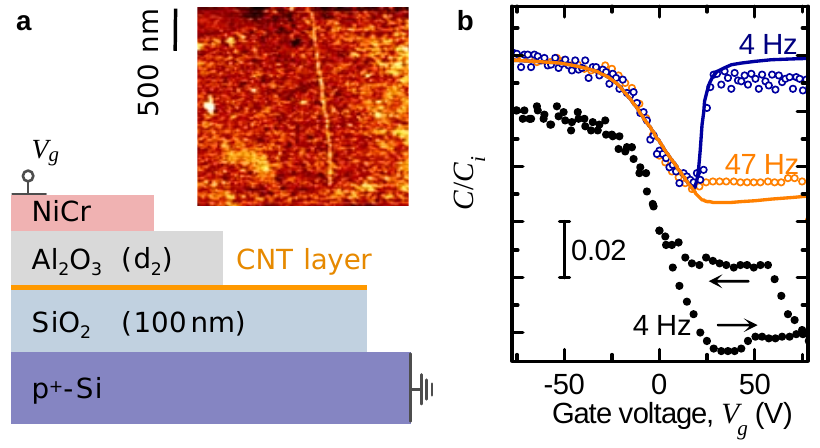}
\centering \caption{(a) Sample schematics: a layer of dispersed
micelle-encapsulated CoMoCat carbon nanotubes with an average
density below $1~\mu$m$^2$ as confirmed by AFM analysis was
embedded in a metal-insulator-semiconductor device with a gate
voltage $V_g$ applied between the semitransparent metallic
nickel-chromium top electrode and $p^+$ doped silicon back gate.
(b) Normalized capacitance $C/C_i$ as a function of the gate
voltage at $4.2$~K with illumination (orange and blue traces) and
without illumination (black traces; arrows represent voltage sweep
directions).} \label{fig1}
\end{figure}

The details of charge traps were investigated with CV spectroscopy
with in situ illumination. In the presence of photo-generated
electrons by diffuse illumination, inversion was recovered for
low-frequency modulation (blue circles in Fig.~\ref{fig1}b) as
opposed to high-frequency modulation (orange circles in
Fig.~\ref{fig1}b). From modeling of the CV characteristics in the
limiting case of low (high) modulation frequency \cite{Grove}
shown as blue (orange) solid line in Fig.~\ref{fig1}c we
determined the characteristic charge impurity density of
$Q_{\textnormal{tot}}=3.6 \cdot 10^{12}~\m{cm}^{-2}$ in our
devices. It includes both the surface states at the $\m{Si-SiO}_2$
interface and the charge states in the insulating oxide volume
\cite{Sze}. Devices of different oxide thicknesses were used to
determine the volume density of the oxide states, $Q_\m{oxide}/d =
3.0 \cdot 10^{17}~\m{cm}^{-3}$. On average this number implies the
presence of a charge trap state within the volume of a cylinder
with $\sim 2$~nm radius around a $200$~nm long CNT. These charge
traps likely constitute the charge reservoir for photoactivated
charge doping of nanotubes that exhibited trion emission.

\clearpage
 {\bf Stark spectroscopy and trion photoluminescence}

\begin{figure}[h!]
\includegraphics[scale=1.0]{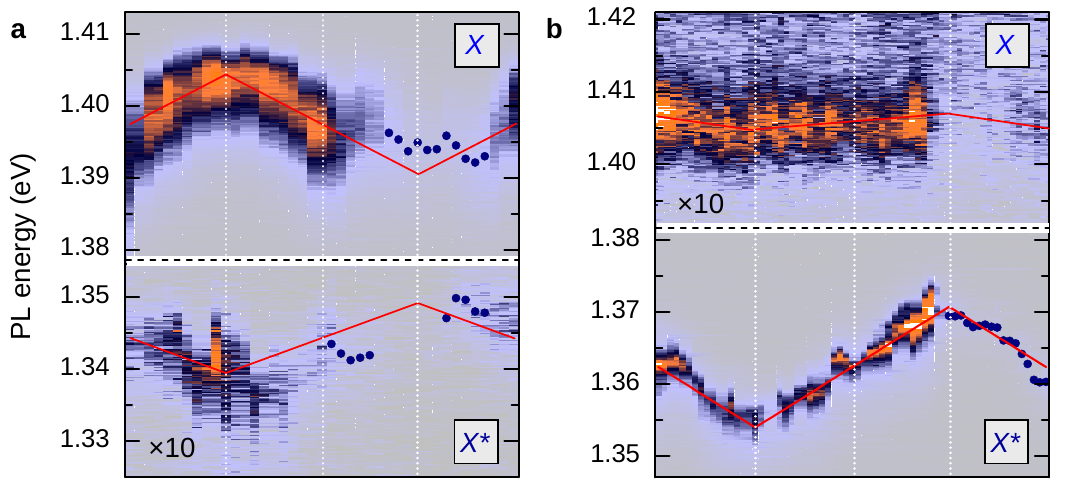}
\vspace{-5pt} \centering \caption{False-color plots of the
photoluminescence from individual nanotubes with double-peak
emission in response to the gate voltage ramp as in Fig.~2a of the
main text: (a) data reproduced from Fig.~2c of the main text, (b)
another nanotube with a double-peak emission and different linear
slopes (red solid lines) associated with the respective permanent
dipole moments $p$. In the regions of low intensity the peak
maxima are shown as blue circles where peak fitting procedure
converged (note also the magnification factors of 10 for the
intensity color-scale in the lower and upper panels of (a) and
(b), respectively).} \label{figSI_E_sweep}
\end{figure}

\begin{figure}[h!]
\includegraphics[scale=1.0]{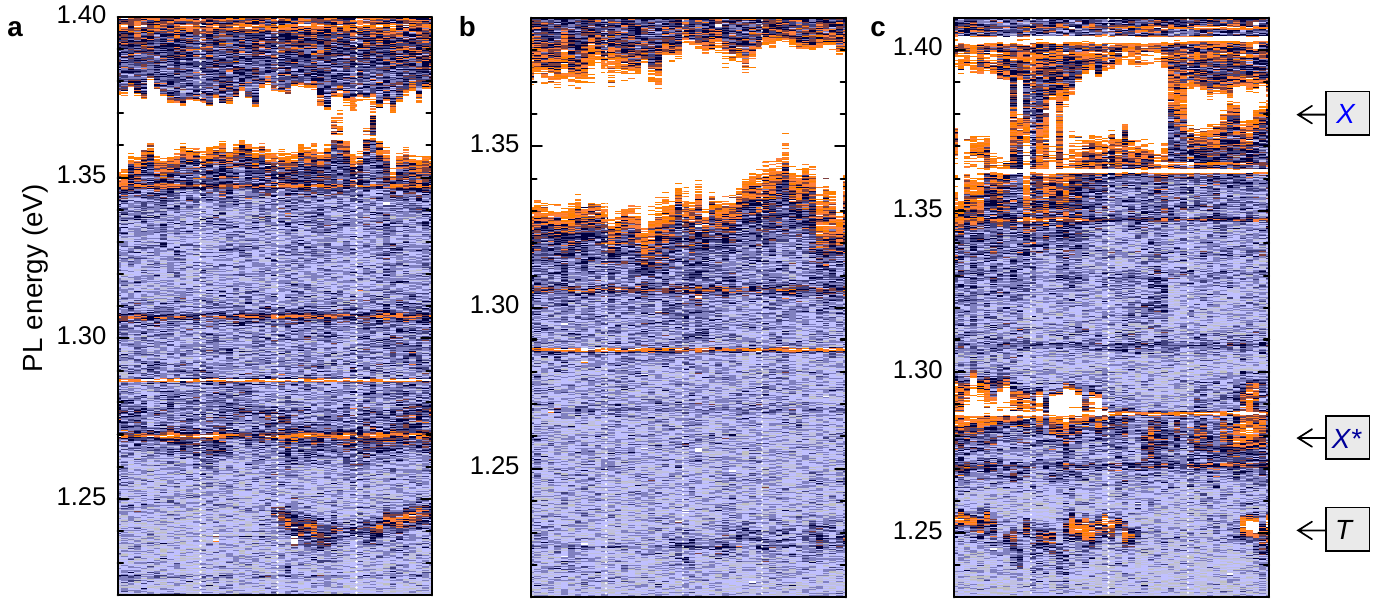}
\vspace{-5pt} \centering \caption{Logarithmic false-color plot of
the photoluminescence dispersion for nanotubes with two emission
peaks ($X$ and $X^*$) in response to the gate voltage ramp as in
Fig.~2a of the main text: (a) data reproduced from Fig.~4 of the
main text, (b) another nanotube with trion emission satellite (T)
at negative gate voltages, and (c) a nanotube with trion emission
at positive gate voltages.} \label{figSI_E_trion}
\end{figure}

Photoluminescence (PL) Stark spectroscopy was performed in
response to transverse electric field as detailed in the main text
and the Methods section. Individual nanotubes exhibited distinct
PL energy shifts according to dipole moments $p$ of different
defect configurations and geometries. Fig.~\ref{figSI_E_sweep}
highlights the case of different responses to a gate voltage ramp
as in Fig.~2a of the main text. The data in
Fig.~\ref{figSI_E_sweep}a is reproduced from Fig.~2c of the main
text for direct comparison with another nanotube shown in
Fig.~\ref{figSI_E_sweep}b with a double-peak spectrum yet
different linear Stark shifts of the PL energy. In
Fig.~\ref{figSI_E_trion} we exemplify the response of different
individual CNTs to charge doping. Two nanotubes exhibited trion
emission satellites at positive gate voltages
(Fig.~\ref{figSI_E_trion}a and b) in contrast to a nanotube with
similar emission characteristics at zero volts and stable trion
emission at negative gate voltages (Fig.~\ref{figSI_E_trion}c).

\clearpage

{\bf Quantum chemistry calculations}

The computations were performed using Gaussian09 software suite
\cite{Frisch2009} with B3LYP functional \cite{Becke1988} and
STO-3G basis set. The dielectric environment due to silicon oxide
and aluminum oxide surroundings of the nanotubes was taken into
account as solvent with a dielectric constant of
$\varepsilon_r=6.3$, which is the average dielectric constant of
the two oxides. The solvent effects were simulated in the
framework of continuum polarizable conductor-like medium
\cite{Cossi2003,Barone1998}; $8$-nm long segments of (6,5) carbon
nanotube with hydrogen-terminated ends were used in these
calculations. Pristine and oxygen-doped CNTs were geometry
optimized in solvent. The optical transition energies were
calculated using Time-Dependent Density Functional Theory
(TD-DFT).

\begin{figure}[h]
\includegraphics[scale=0.9]{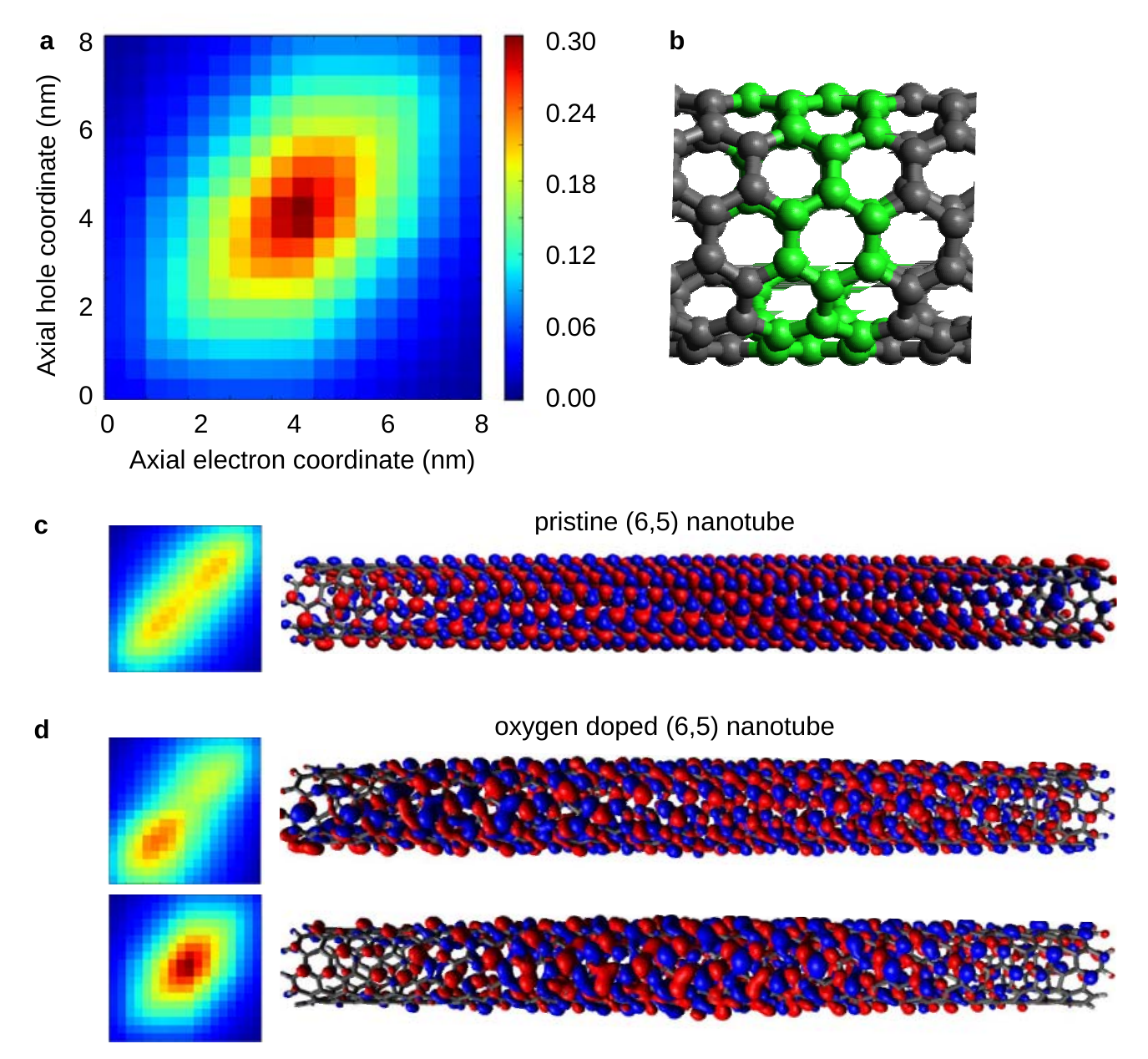}
\centering \caption{(a) Contour plot of the transition density of
the excitonic state which corresponds to the $E_{11}^{*}$ level of
the doped tube. Horizontal and vertical axes show the distribution
of electron and hole densities, respectively. (b) The size of
pixel in contour plot corresponds to the section of the nanotube
highlighted in green. It contains $39$ carbon atoms and its width
is $4.4$~\AA. (c) Exciton contour plot and orbital projection of
transition density for $E_{11}$ level of a (6,5) pristine
nanotube. (d) Exciton contour plots and orbital projection of
transition density of a (6,5) oxygen-doped nanotube.}
\label{figSI_T_1}
\end{figure}

Electric field was applied in transverse direction and optical
transition energies were computed without additional geometry
optimization. The transition energies, shifted by the energy of
$E_{11}$ transition in pristine CNT, are shown in
Fig.~\ref{figSI_T_disp}. In order to establish a quantitative
measure of exciton wave functions, exciton plots were built for
relevant optical transitions and shown in the left panels of
Fig.s~\ref{figSI_T_1}, \ref{figSI_T_1} and \ref{figSI_T_1} (see
Ref.s~\cite{Mukamel1997,Tretiak2002} for more details on exciton
plot characterization of excited states in one-dimensional
structures). In these contour plots, the bright spot elongated
along the diagonal signifies where the exciton wave function is
located along the CNT axis, and the "width" of the elongated plot
is the electron-hole correlation length (also referred to as the
exciton size). Our analysis shows that excitons are localized on
sub-$10$~nm length scale (to about $3$~nm for the $E_{11}^{*}$
state, denoted as $X^*$ in the main text, and about $6$~nm for the
$E_{11}^{-}$ state denoted as $X$ in the main text) with an
electron-hole correlation length of about $2$~nm. We also
projected the exciton wave function onto the basis of atomic
orbitals in order to provide a qualitative real-space visual
measure of exciton wave function. The corresponding plots are
shown in the right panels of Fig.s~\ref{figSI_T_1},
\ref{figSI_T_1} and \ref{figSI_T_1}.

\begin{figure}[ht!]
\includegraphics[scale=0.88]{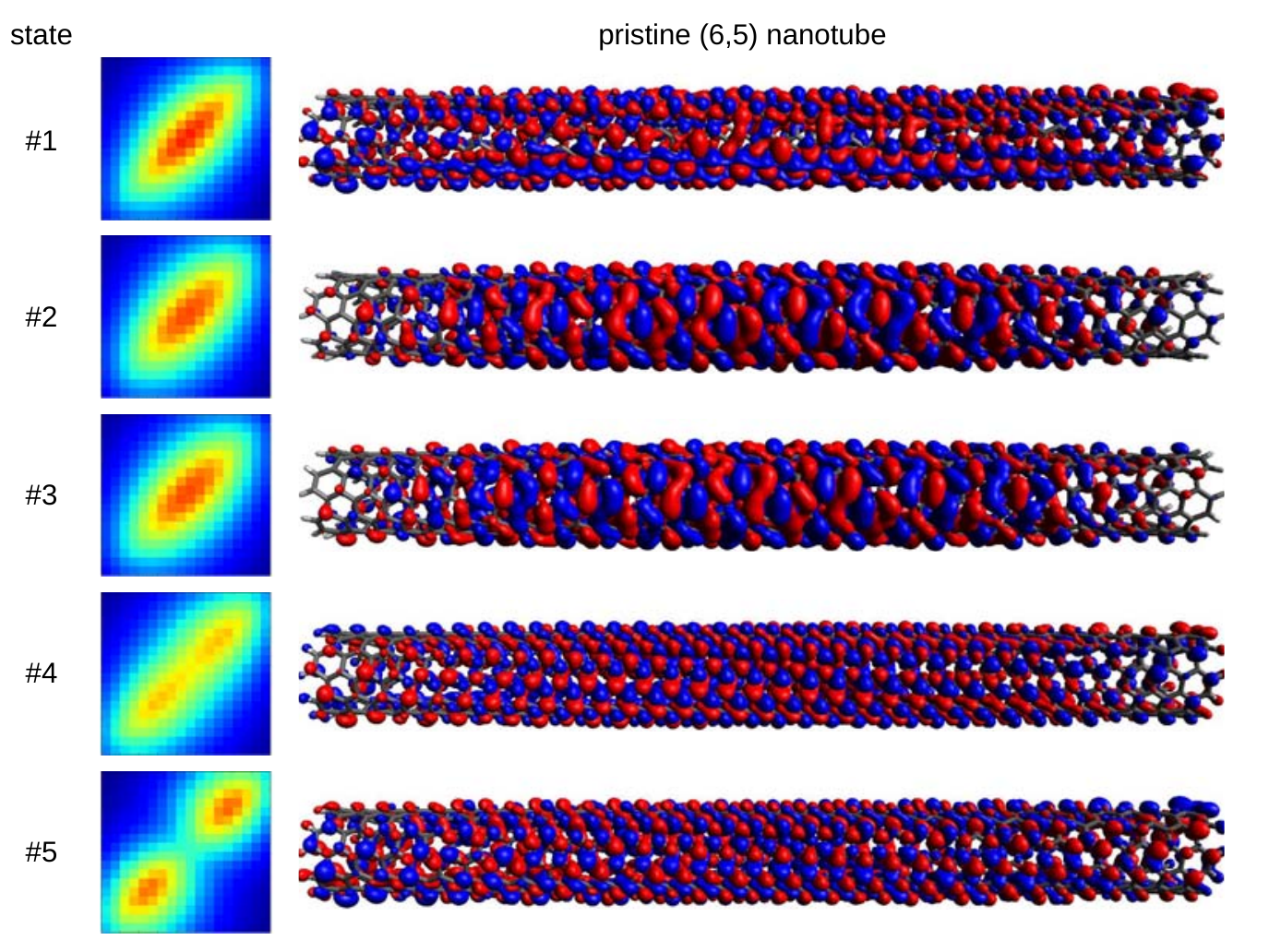}
\vspace{-10pt} \centering \caption{Exciton plots of the five
lowest optical transitions of a pristine (6,5) nanotube in a
dielectric medium with $\varepsilon_r=6.3$. States 1-3 are dark
excitons and state 4 is the bright $E_{11}$ exciton. State 5,
consisting of two lobes, is the higher excited state that is
artificially brightened on the ends of the nanotube.}
\label{figSI_T_2}
\end{figure}


\begin{figure}[hb!]
\includegraphics[scale=0.88]{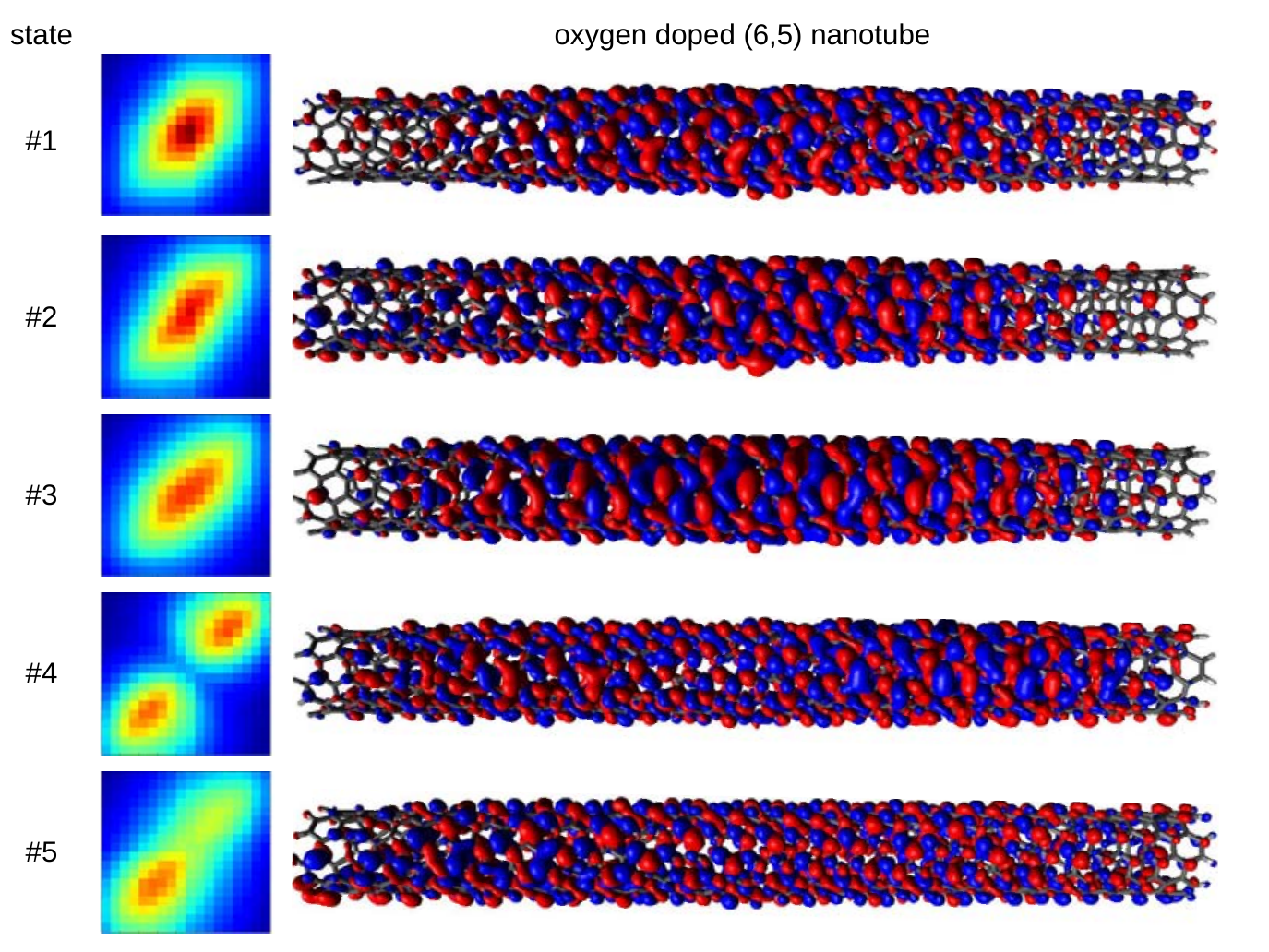}
\centering \caption{Exciton plots for the lowest
five optical transitions in an oxygen-doped (6,5) nanotube in a
dielectric medium with $\varepsilon_r=6.3$. The dopant is located
in the center of a $8$-nm long nanotube segment. State 1 is the
localized bright exciton $E_{11}^{*}$, denoted as $X^*$ in the
main text. States 1 and 2 are quasi-degenerate and both localized.
State 4 is the higher excited state, which was excluded from
theoretical analysis because of an unjustified value of its
oscillator strength due to artificial brightening at the ends of
the nanotube. State 5, denoted as $X$ in the main text, is
$E_{11}^{-}$ exciton delocalized but substantially perturbed by
the defect.} \label{figSI_T_3}
\end{figure}


\begin{figure}[h]
\includegraphics[scale=0.97]{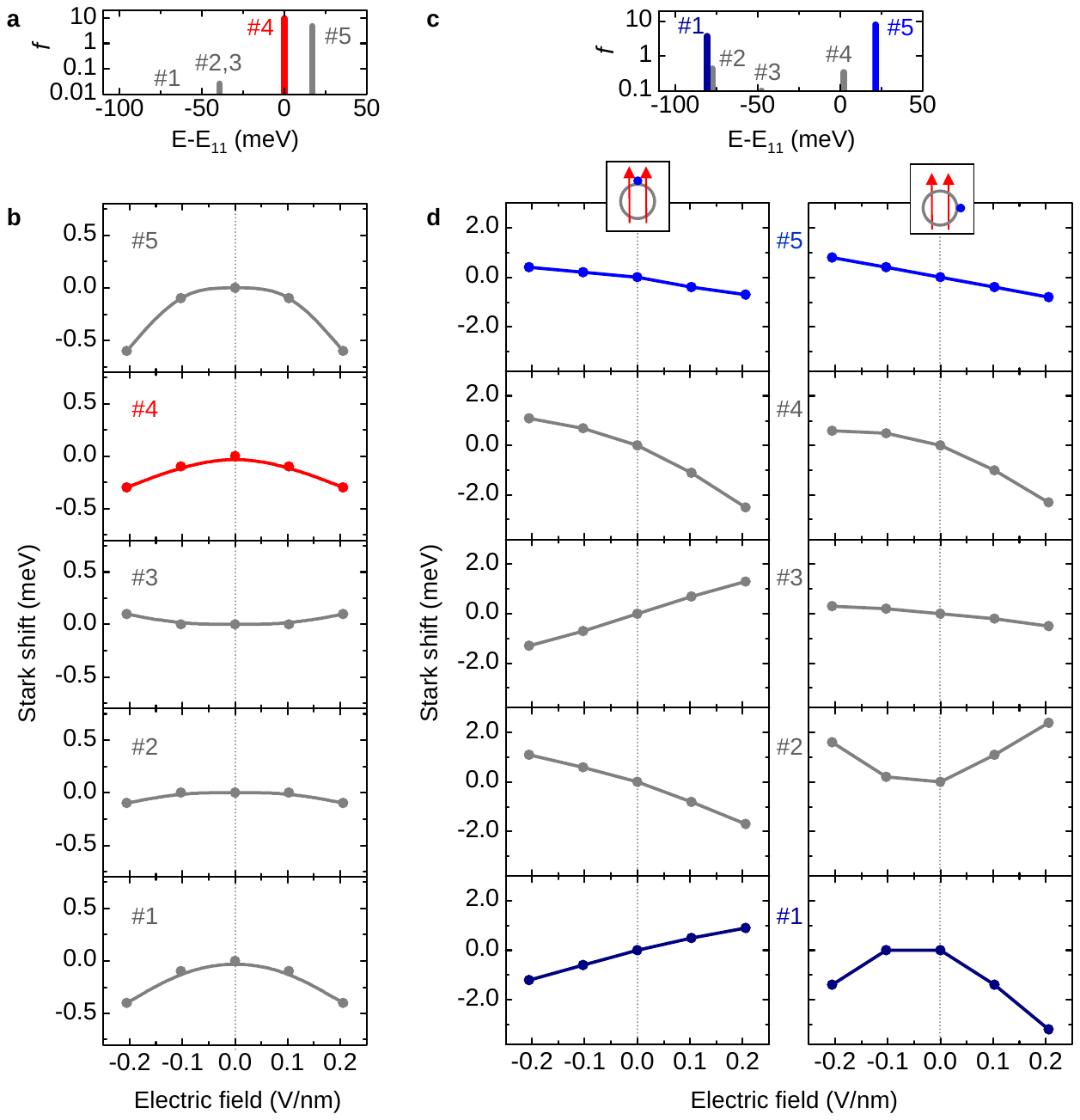}
\centering \caption{(a) Calculated optical spectrum
of a pristine (6,5) nanotube in a dielectric medium with
$\varepsilon_r=6.3$ (the corresponding exciton plots of states 1
to 5 are shown in Fig.~\ref{figSI_T_2}). (b) Energy dispersion of
the respective states in transverse electric field. (c) Calculated
optical spectrum of an oxygen-doped (6,5) nanotube in a dielectric
medium with $\varepsilon_r=6.3$ (the corresponding exciton plots
of states 1 to 5 are shown in Fig.~\ref{figSI_T_3}). (d) Energy
dispersion of the respective defect states in transverse electric
field for two different defect configurations (left panel: defect
on the apex, right panel: peripheral defect).}
\label{figSI_T_disp}
\end{figure}


\begin{thebibliography}{50}
\expandafter\ifx\csname
natexlab\endcsname\relax\def\natexlab#1{#1}\fi
\expandafter\ifx\csname bibnamefont\endcsname\relax
  \def\bibnamefont#1{#1}\fi
\expandafter\ifx\csname bibfnamefont\endcsname\relax
  \def\bibfnamefont#1{#1}\fi
\expandafter\ifx\csname citenamefont\endcsname\relax
  \def\citenamefont#1{#1}\fi
\expandafter\ifx\csname url\endcsname\relax
  \def\url#1{\texttt{#1}}\fi
\expandafter\ifx\csname
urlprefix\endcsname\relax\def\urlprefix{URL }\fi
\providecommand{\bibinfo}[2]{#2}
\providecommand{\eprint}[2][]{\url{#2}}

\bibitem[{\citenamefont{Wang et~al.}(2005)\citenamefont{Wang, Dukovic, Brus,
  and Heinz}}]{Wang2005}
\bibinfo{author}{\bibfnamefont{F.}~\bibnamefont{Wang}},
  \bibinfo{author}{\bibfnamefont{G.}~\bibnamefont{Dukovic}},
  \bibinfo{author}{\bibfnamefont{L.~E.} \bibnamefont{Brus}}, \bibnamefont{and}
  \bibinfo{author}{\bibfnamefont{T.~F.} \bibnamefont{Heinz}},
  \bibinfo{journal}{Science} \textbf{\bibinfo{volume}{308}},
  \bibinfo{pages}{838} (\bibinfo{year}{2005}).

\bibitem[{\citenamefont{Maultzsch et~al.}(2005)\citenamefont{Maultzsch,
  Pomraenke, Reich, Chang, Prezzi, Ruini, Molinari, Strano, Thomsen, and
  Lienau}}]{Maultzsch2005}
\bibinfo{author}{\bibfnamefont{J.}~\bibnamefont{Maultzsch}},
  \bibinfo{author}{\bibfnamefont{R.}~\bibnamefont{Pomraenke}},
  \bibinfo{author}{\bibfnamefont{S.}~\bibnamefont{Reich}},
  \bibinfo{author}{\bibfnamefont{E.}~\bibnamefont{Chang}},
  \bibinfo{author}{\bibfnamefont{D.}~\bibnamefont{Prezzi}},
  \bibinfo{author}{\bibfnamefont{A.}~\bibnamefont{Ruini}},
  \bibinfo{author}{\bibfnamefont{E.}~\bibnamefont{Molinari}},
  \bibinfo{author}{\bibfnamefont{M.~S.} \bibnamefont{Strano}},
  \bibinfo{author}{\bibfnamefont{C.}~\bibnamefont{Thomsen}}, \bibnamefont{and}
  \bibinfo{author}{\bibfnamefont{C.}~\bibnamefont{Lienau}},
  \bibinfo{journal}{Phys. Rev. B} \textbf{\bibinfo{volume}{72}},
  \bibinfo{pages}{241402} (\bibinfo{year}{2005}).

\bibitem[{\citenamefont{H\"ogele et~al.}(2008)\citenamefont{H\"ogele, Galland,
  Winger, and Imamoglu}}]{Hogele2008}
\bibinfo{author}{\bibfnamefont{A.}~\bibnamefont{H\"ogele}},
  \bibinfo{author}{\bibfnamefont{C.}~\bibnamefont{Galland}},
  \bibinfo{author}{\bibfnamefont{M.}~\bibnamefont{Winger}}, \bibnamefont{and}
  \bibinfo{author}{\bibfnamefont{A.}~\bibnamefont{Imamoglu}},
  \bibinfo{journal}{Phys. Rev. Lett.} \textbf{\bibinfo{volume}{100}},
  \bibinfo{pages}{217401} (\bibinfo{year}{2008}).

\bibitem[{\citenamefont{Georgi et~al.}(2010)\citenamefont{Georgi, Green,
  Hersam, and Hartschuh}}]{Georgi2010}
\bibinfo{author}{\bibfnamefont{C.}~\bibnamefont{Georgi}},
  \bibinfo{author}{\bibfnamefont{A.~A.} \bibnamefont{Green}},
  \bibinfo{author}{\bibfnamefont{M.~C.} \bibnamefont{Hersam}},
  \bibnamefont{and}
  \bibinfo{author}{\bibfnamefont{A.}~\bibnamefont{Hartschuh}},
  \bibinfo{journal}{ACS Nano} \textbf{\bibinfo{volume}{4}},
  \bibinfo{pages}{5914} (\bibinfo{year}{2010}).

\bibitem[{\citenamefont{Hofmann et~al.}(2016)\citenamefont{Hofmann, Noe, Kneer,
  Crochet, and H\"ogele}}]{Hofmann2016}
\bibinfo{author}{\bibfnamefont{M.~S.} \bibnamefont{Hofmann}},
  \bibinfo{author}{\bibfnamefont{J.}~\bibnamefont{Noe}},
  \bibinfo{author}{\bibfnamefont{A.}~\bibnamefont{Kneer}},
  \bibinfo{author}{\bibfnamefont{J.~J.} \bibnamefont{Crochet}},
  \bibnamefont{and} \bibinfo{author}{\bibfnamefont{A.}~\bibnamefont{H\"ogele}},
  \bibinfo{journal}{Nano Lett.} \textbf{\bibinfo{volume}{16}},
  \bibinfo{pages}{2958} (\bibinfo{year}{2016}).

\bibitem[{\citenamefont{Tayo and Rotkin}(2012)}]{Tayo2012}
\bibinfo{author}{\bibfnamefont{B.~O.} \bibnamefont{Tayo}} \bibnamefont{and}
  \bibinfo{author}{\bibfnamefont{S.~V.} \bibnamefont{Rotkin}},
  \bibinfo{journal}{Phys. Rev. B} \textbf{\bibinfo{volume}{86}},
  \bibinfo{pages}{125431} (\bibinfo{year}{2012}).

\bibitem[{\citenamefont{Ma et~al.}(2015{\natexlab{a}})\citenamefont{Ma,
  Hartmann, Baldwin, Doorn, and Htoon}}]{Ma2015nn}
\bibinfo{author}{\bibfnamefont{X.}~\bibnamefont{Ma}},
  \bibinfo{author}{\bibfnamefont{N.~F.} \bibnamefont{Hartmann}},
  \bibinfo{author}{\bibfnamefont{J.~K.~S.} \bibnamefont{Baldwin}},
  \bibinfo{author}{\bibfnamefont{S.~K.} \bibnamefont{Doorn}}, \bibnamefont{and}
  \bibinfo{author}{\bibfnamefont{H.}~\bibnamefont{Htoon}},
  \bibinfo{journal}{Nat. Nanotechnol.} \textbf{\bibinfo{volume}{10}},
  \bibinfo{pages}{671} (\bibinfo{year}{2015}{\natexlab{a}}).

\bibitem[{\citenamefont{Ghosh et~al.}(2010)\citenamefont{Ghosh, Bachilo,
  Simonette, Beckingham, and Weisman}}]{Ghosh2010}
\bibinfo{author}{\bibfnamefont{S.}~\bibnamefont{Ghosh}},
  \bibinfo{author}{\bibfnamefont{S.~M.} \bibnamefont{Bachilo}},
  \bibinfo{author}{\bibfnamefont{R.~A.} \bibnamefont{Simonette}},
  \bibinfo{author}{\bibfnamefont{K.~M.} \bibnamefont{Beckingham}},
  \bibnamefont{and} \bibinfo{author}{\bibfnamefont{R.~B.}
  \bibnamefont{Weisman}}, \bibinfo{journal}{Science}
  \textbf{\bibinfo{volume}{330}}, \bibinfo{pages}{1656} (\bibinfo{year}{2010}).

\bibitem[{\citenamefont{Ma et~al.}(2014{\natexlab{a}})\citenamefont{Ma,
  Adamska, Yamaguchi, Yalcin, Tretiak, Doorn, and Htoon}}]{Ma2014a}
\bibinfo{author}{\bibfnamefont{X.}~\bibnamefont{Ma}},
  \bibinfo{author}{\bibfnamefont{L.}~\bibnamefont{Adamska}},
  \bibinfo{author}{\bibfnamefont{H.}~\bibnamefont{Yamaguchi}},
  \bibinfo{author}{\bibfnamefont{S.~E.} \bibnamefont{Yalcin}},
  \bibinfo{author}{\bibfnamefont{S.}~\bibnamefont{Tretiak}},
  \bibinfo{author}{\bibfnamefont{S.~K.} \bibnamefont{Doorn}}, \bibnamefont{and}
  \bibinfo{author}{\bibfnamefont{H.}~\bibnamefont{Htoon}},
  \bibinfo{journal}{ACS Nano} \textbf{\bibinfo{volume}{8}},
  \bibinfo{pages}{10782} (\bibinfo{year}{2014}{\natexlab{a}}).

\bibitem[{\citenamefont{Ma et~al.}(2015{\natexlab{b}})\citenamefont{Ma,
  Baldwin, Hartmann, Doorn, and Htoon}}]{Ma2015adv}
\bibinfo{author}{\bibfnamefont{X.}~\bibnamefont{Ma}},
  \bibinfo{author}{\bibfnamefont{J.~K.} \bibnamefont{Baldwin}},
  \bibinfo{author}{\bibfnamefont{N.~F.} \bibnamefont{Hartmann}},
  \bibinfo{author}{\bibfnamefont{S.~K.} \bibnamefont{Doorn}}, \bibnamefont{and}
  \bibinfo{author}{\bibfnamefont{H.}~\bibnamefont{Htoon}},
  \bibinfo{journal}{Adv. Funct. Mater.} \textbf{\bibinfo{volume}{25}},
  \bibinfo{pages}{6157} (\bibinfo{year}{2015}{\natexlab{b}}).

\bibitem[{\citenamefont{Piao et~al.}(2013)\citenamefont{Piao, Meany, Powell,
  Valley, Kwon, Schatz, and Wang}}]{Piao2013}
\bibinfo{author}{\bibfnamefont{Y.}~\bibnamefont{Piao}},
  \bibinfo{author}{\bibfnamefont{B.}~\bibnamefont{Meany}},
  \bibinfo{author}{\bibfnamefont{L.~R.} \bibnamefont{Powell}},
  \bibinfo{author}{\bibfnamefont{N.}~\bibnamefont{Valley}},
  \bibinfo{author}{\bibfnamefont{H.}~\bibnamefont{Kwon}},
  \bibinfo{author}{\bibfnamefont{G.~C.} \bibnamefont{Schatz}},
  \bibnamefont{and} \bibinfo{author}{\bibfnamefont{Y.}~\bibnamefont{Wang}},
  \bibinfo{journal}{Nat Chem} \textbf{\bibinfo{volume}{5}},
  \bibinfo{pages}{840} (\bibinfo{year}{2013}).

\bibitem[{\citenamefont{Kwon et~al.}(2016)\citenamefont{Kwon, Furmanchuk, Kim,
  Meany, Guo, Schatz, and Wang}}]{Kwon2016}
\bibinfo{author}{\bibfnamefont{H.}~\bibnamefont{Kwon}},
  \bibinfo{author}{\bibfnamefont{A.}~\bibnamefont{Furmanchuk}},
  \bibinfo{author}{\bibfnamefont{M.}~\bibnamefont{Kim}},
  \bibinfo{author}{\bibfnamefont{B.}~\bibnamefont{Meany}},
  \bibinfo{author}{\bibfnamefont{Y.}~\bibnamefont{Guo}},
  \bibinfo{author}{\bibfnamefont{G.~C.} \bibnamefont{Schatz}},
  \bibnamefont{and} \bibinfo{author}{\bibfnamefont{Y.}~\bibnamefont{Wang}},
  \bibinfo{journal}{J. Am. Chem. Soc.} \textbf{\bibinfo{volume}{131}},
  \bibinfo{pages}{6878} (\bibinfo{year}{2016}).

\bibitem[{\citenamefont{Miyauchi et~al.}(2013)\citenamefont{Miyauchi, Iwamura,
  Mouri, Kawazoe, Ohtsu, and Matsuda}}]{Miyauchi2013}
\bibinfo{author}{\bibfnamefont{Y.}~\bibnamefont{Miyauchi}},
  \bibinfo{author}{\bibfnamefont{M.}~\bibnamefont{Iwamura}},
  \bibinfo{author}{\bibfnamefont{S.}~\bibnamefont{Mouri}},
  \bibinfo{author}{\bibfnamefont{T.}~\bibnamefont{Kawazoe}},
  \bibinfo{author}{\bibfnamefont{M.}~\bibnamefont{Ohtsu}}, \bibnamefont{and}
  \bibinfo{author}{\bibfnamefont{K.}~\bibnamefont{Matsuda}},
  \bibinfo{journal}{Nat. Photon.} \textbf{\bibinfo{volume}{7}},
  \bibinfo{pages}{715} (\bibinfo{year}{2013}).

\bibitem[{\citenamefont{Hartmann et~al.}(2015)\citenamefont{Hartmann, Yalcin,
  Adamska, H{\'a}roz, Ma, Tretiak, Htoon, and Doorn}}]{Hartmann2015}
\bibinfo{author}{\bibfnamefont{N.~F.} \bibnamefont{Hartmann}},
  \bibinfo{author}{\bibfnamefont{S.~E.} \bibnamefont{Yalcin}},
  \bibinfo{author}{\bibfnamefont{L.}~\bibnamefont{Adamska}},
  \bibinfo{author}{\bibfnamefont{E.~H.} \bibnamefont{H{\'a}roz}},
  \bibinfo{author}{\bibfnamefont{X.}~\bibnamefont{Ma}},
  \bibinfo{author}{\bibfnamefont{S.}~\bibnamefont{Tretiak}},
  \bibinfo{author}{\bibfnamefont{H.}~\bibnamefont{Htoon}}, \bibnamefont{and}
  \bibinfo{author}{\bibfnamefont{S.~K.} \bibnamefont{Doorn}},
  \bibinfo{journal}{Nanoscale} \textbf{\bibinfo{volume}{7}},
  \bibinfo{pages}{20521} (\bibinfo{year}{2015}).

\bibitem[{\citenamefont{Ma et~al.}(2014{\natexlab{b}})\citenamefont{Ma,
  Roslyak, Wang, Duque, Piryatinski, Doorn, and Htoon}}]{Ma2014b}
\bibinfo{author}{\bibfnamefont{X.}~\bibnamefont{Ma}},
  \bibinfo{author}{\bibfnamefont{O.}~\bibnamefont{Roslyak}},
  \bibinfo{author}{\bibfnamefont{F.}~\bibnamefont{Wang}},
  \bibinfo{author}{\bibfnamefont{J.~G.} \bibnamefont{Duque}},
  \bibinfo{author}{\bibfnamefont{A.}~\bibnamefont{Piryatinski}},
  \bibinfo{author}{\bibfnamefont{S.~K.} \bibnamefont{Doorn}}, \bibnamefont{and}
  \bibinfo{author}{\bibfnamefont{H.}~\bibnamefont{Htoon}},
  \bibinfo{journal}{ACS Nano} \textbf{\bibinfo{volume}{8}},
  \bibinfo{pages}{10613} (\bibinfo{year}{2014}{\natexlab{b}}).

\bibitem[{\citenamefont{Ma et~al.}(2015{\natexlab{c}})\citenamefont{Ma,
  Roslyak, Duque, Pang, Doorn, Piryatinski, Dunlap, and Htoon}}]{Ma2015prl}
\bibinfo{author}{\bibfnamefont{X.}~\bibnamefont{Ma}},
  \bibinfo{author}{\bibfnamefont{O.}~\bibnamefont{Roslyak}},
  \bibinfo{author}{\bibfnamefont{J.~G.} \bibnamefont{Duque}},
  \bibinfo{author}{\bibfnamefont{X.}~\bibnamefont{Pang}},
  \bibinfo{author}{\bibfnamefont{S.~K.} \bibnamefont{Doorn}},
  \bibinfo{author}{\bibfnamefont{A.}~\bibnamefont{Piryatinski}},
  \bibinfo{author}{\bibfnamefont{D.~H.} \bibnamefont{Dunlap}},
  \bibnamefont{and} \bibinfo{author}{\bibfnamefont{H.}~\bibnamefont{Htoon}},
  \bibinfo{journal}{Phys. Rev. Lett.} \textbf{\bibinfo{volume}{115}},
  \bibinfo{pages}{017401} (\bibinfo{year}{2015}{\natexlab{c}}).

\bibitem[{\citenamefont{Hartmann et~al.}(2016)\citenamefont{Hartmann,
  Velizhanin, Haroz, Kim, Ma, Wang, Htoon, and Doorn}}]{Hartmann2016}
\bibinfo{author}{\bibfnamefont{N.~F.} \bibnamefont{Hartmann}},
  \bibinfo{author}{\bibfnamefont{K.~A.} \bibnamefont{Velizhanin}},
  \bibinfo{author}{\bibfnamefont{E.~H.} \bibnamefont{Haroz}},
  \bibinfo{author}{\bibfnamefont{M.}~\bibnamefont{Kim}},
  \bibinfo{author}{\bibfnamefont{X.}~\bibnamefont{Ma}},
  \bibinfo{author}{\bibfnamefont{Y.}~\bibnamefont{Wang}},
  \bibinfo{author}{\bibfnamefont{H.}~\bibnamefont{Htoon}}, \bibnamefont{and}
  \bibinfo{author}{\bibfnamefont{S.~K.} \bibnamefont{Doorn}},
  \bibinfo{journal}{ACS nano} \textbf{\bibinfo{volume}{10}},
  \bibinfo{pages}{8355} (\bibinfo{year}{2016}).

\bibitem[{\citenamefont{Jeantet et~al.}(2016)\citenamefont{Jeantet,
  Chassagneux, Raynaud, Roussignol, Lauret, Besga, Est\`eve, Reichel, and
  Voisin}}]{Jeantet2016}
\bibinfo{author}{\bibfnamefont{A.}~\bibnamefont{Jeantet}},
  \bibinfo{author}{\bibfnamefont{Y.}~\bibnamefont{Chassagneux}},
  \bibinfo{author}{\bibfnamefont{C.}~\bibnamefont{Raynaud}},
  \bibinfo{author}{\bibfnamefont{P.}~\bibnamefont{Roussignol}},
  \bibinfo{author}{\bibfnamefont{J.~S.} \bibnamefont{Lauret}},
  \bibinfo{author}{\bibfnamefont{B.}~\bibnamefont{Besga}},
  \bibinfo{author}{\bibfnamefont{J.}~\bibnamefont{Est\`eve}},
  \bibinfo{author}{\bibfnamefont{J.}~\bibnamefont{Reichel}}, \bibnamefont{and}
  \bibinfo{author}{\bibfnamefont{C.}~\bibnamefont{Voisin}},
  \bibinfo{journal}{Phys. Rev. Lett.} \textbf{\bibinfo{volume}{116}},
  \bibinfo{pages}{247402} (\bibinfo{year}{2016}).

\bibitem[{\citenamefont{H\"ummer et~al.}(2016)\citenamefont{H\"ummer, Noe,
  Hofmann, H\"ansch, H\"ogele, and Hunger}}]{Hummer2016}
\bibinfo{author}{\bibfnamefont{T.}~\bibnamefont{H\"ummer}},
  \bibinfo{author}{\bibfnamefont{J.}~\bibnamefont{Noe}},
  \bibinfo{author}{\bibfnamefont{M.~S.} \bibnamefont{Hofmann}},
  \bibinfo{author}{\bibfnamefont{T.~W.} \bibnamefont{H\"ansch}},
  \bibinfo{author}{\bibfnamefont{A.}~\bibnamefont{H\"ogele}}, \bibnamefont{and}
  \bibinfo{author}{\bibfnamefont{D.}~\bibnamefont{Hunger}},
  \bibinfo{journal}{Nature Communications} \textbf{\bibinfo{volume}{7}},
  \bibinfo{pages}{12155} (\bibinfo{year}{2016}).

\bibitem[{\citenamefont{Matsunaga et~al.}(2011)\citenamefont{Matsunaga,
  Matsuda, and Kanemitsu}}]{Matsunaga2011}
\bibinfo{author}{\bibfnamefont{R.}~\bibnamefont{Matsunaga}},
  \bibinfo{author}{\bibfnamefont{K.}~\bibnamefont{Matsuda}}, \bibnamefont{and}
  \bibinfo{author}{\bibfnamefont{Y.}~\bibnamefont{Kanemitsu}},
  \bibinfo{journal}{Phys. Rev. Lett.} \textbf{\bibinfo{volume}{106}},
  \bibinfo{pages}{037404} (\bibinfo{year}{2011}).

\bibitem[{\citenamefont{Park et~al.}(2012)\citenamefont{Park, Hirana, Mouri,
  Miyauchi, Nakashima, and Matsuda}}]{Park2012}
\bibinfo{author}{\bibfnamefont{J.~S.} \bibnamefont{Park}},
  \bibinfo{author}{\bibfnamefont{Y.}~\bibnamefont{Hirana}},
  \bibinfo{author}{\bibfnamefont{S.}~\bibnamefont{Mouri}},
  \bibinfo{author}{\bibfnamefont{Y.}~\bibnamefont{Miyauchi}},
  \bibinfo{author}{\bibfnamefont{N.}~\bibnamefont{Nakashima}},
  \bibnamefont{and} \bibinfo{author}{\bibfnamefont{K.}~\bibnamefont{Matsuda}},
  \bibinfo{journal}{J. Am. Chem. Soc.} \textbf{\bibinfo{volume}{134}},
  \bibinfo{pages}{14461} (\bibinfo{year}{2012}).

\bibitem[{\citenamefont{Brozena et~al.}(2014)\citenamefont{Brozena, Leeds,
  Zhang, Fourkas, and Wang}}]{Brozena2014}
\bibinfo{author}{\bibfnamefont{A.~H.} \bibnamefont{Brozena}},
  \bibinfo{author}{\bibfnamefont{J.~D.} \bibnamefont{Leeds}},
  \bibinfo{author}{\bibfnamefont{Y.}~\bibnamefont{Zhang}},
  \bibinfo{author}{\bibfnamefont{J.~T.} \bibnamefont{Fourkas}},
  \bibnamefont{and} \bibinfo{author}{\bibfnamefont{Y.}~\bibnamefont{Wang}},
  \bibinfo{journal}{ACS Nano} \textbf{\bibinfo{volume}{8}},
  \bibinfo{pages}{4239} (\bibinfo{year}{2014}).

\bibitem[{\citenamefont{Galland and Imamoglu}(2008)}]{Galland2008spin}
\bibinfo{author}{\bibfnamefont{C.}~\bibnamefont{Galland}} \bibnamefont{and}
  \bibinfo{author}{\bibfnamefont{A.}~\bibnamefont{Imamoglu}},
  \bibinfo{journal}{Phys. Rev. Lett.} \textbf{\bibinfo{volume}{101}},
  \bibinfo{pages}{157404} (\bibinfo{year}{2008}).

\bibitem[{\citenamefont{Gao et~al.}(2015)\citenamefont{Gao, Imamoglu, Bernien,
  and Hanson}}]{Gao2015}
\bibinfo{author}{\bibfnamefont{W.~B.} \bibnamefont{Gao}},
  \bibinfo{author}{\bibfnamefont{A.}~\bibnamefont{Imamoglu}},
  \bibinfo{author}{\bibfnamefont{H.}~\bibnamefont{Bernien}}, \bibnamefont{and}
  \bibinfo{author}{\bibfnamefont{R.}~\bibnamefont{Hanson}},
  \bibinfo{journal}{Nat Photon} \textbf{\bibinfo{volume}{9}},
  \bibinfo{pages}{363} (\bibinfo{year}{2015}).

\bibitem[{\citenamefont{Maze et~al.}(2008)\citenamefont{Maze, Stanwix, Hodges,
  Hong, Taylor, Cappellaro, Jiang, Dutt, Togan, Zibrov et~al.}}]{Maze2008}
\bibinfo{author}{\bibfnamefont{J.~R.} \bibnamefont{Maze}},
  \bibinfo{author}{\bibfnamefont{P.~L.} \bibnamefont{Stanwix}},
  \bibinfo{author}{\bibfnamefont{J.~S.} \bibnamefont{Hodges}},
  \bibinfo{author}{\bibfnamefont{S.}~\bibnamefont{Hong}},
  \bibinfo{author}{\bibfnamefont{J.~M.} \bibnamefont{Taylor}},
  \bibinfo{author}{\bibfnamefont{P.}~\bibnamefont{Cappellaro}},
  \bibinfo{author}{\bibfnamefont{L.}~\bibnamefont{Jiang}},
  \bibinfo{author}{\bibfnamefont{M.~V.~G.} \bibnamefont{Dutt}},
  \bibinfo{author}{\bibfnamefont{E.}~\bibnamefont{Togan}},
  \bibinfo{author}{\bibfnamefont{A.~S.} \bibnamefont{Zibrov}},
  \bibnamefont{et~al.}, \bibinfo{journal}{Nature}
  \textbf{\bibinfo{volume}{455}}, \bibinfo{pages}{644} (\bibinfo{year}{2008}).

\bibitem[{\citenamefont{Balasubramanian
  et~al.}(2008)\citenamefont{Balasubramanian, Chan, Kolesov, Al-Hmoud, Tisler,
  Shin, Kim, Wojcik, Hemmer, Krueger et~al.}}]{Balasubramanian2008}
\bibinfo{author}{\bibfnamefont{G.}~\bibnamefont{Balasubramanian}},
  \bibinfo{author}{\bibfnamefont{I.~Y.} \bibnamefont{Chan}},
  \bibinfo{author}{\bibfnamefont{R.}~\bibnamefont{Kolesov}},
  \bibinfo{author}{\bibfnamefont{M.}~\bibnamefont{Al-Hmoud}},
  \bibinfo{author}{\bibfnamefont{J.}~\bibnamefont{Tisler}},
  \bibinfo{author}{\bibfnamefont{C.}~\bibnamefont{Shin}},
  \bibinfo{author}{\bibfnamefont{C.}~\bibnamefont{Kim}},
  \bibinfo{author}{\bibfnamefont{A.}~\bibnamefont{Wojcik}},
  \bibinfo{author}{\bibfnamefont{P.~R.} \bibnamefont{Hemmer}},
  \bibinfo{author}{\bibfnamefont{A.}~\bibnamefont{Krueger}},
  \bibnamefont{et~al.}, \bibinfo{journal}{Nature}
  \textbf{\bibinfo{volume}{455}}, \bibinfo{pages}{648} (\bibinfo{year}{2008}).

\bibitem[{\citenamefont{Vamivakas et~al.}(2011)\citenamefont{Vamivakas, Zhao,
  F\"alt, Badolato, Taylor, and Atat\"ure}}]{Vamivakas2011}
\bibinfo{author}{\bibfnamefont{A.~N.} \bibnamefont{Vamivakas}},
  \bibinfo{author}{\bibfnamefont{Y.}~\bibnamefont{Zhao}},
  \bibinfo{author}{\bibfnamefont{S.}~\bibnamefont{F\"alt}},
  \bibinfo{author}{\bibfnamefont{A.}~\bibnamefont{Badolato}},
  \bibinfo{author}{\bibfnamefont{J.~M.} \bibnamefont{Taylor}},
  \bibnamefont{and}
  \bibinfo{author}{\bibfnamefont{M.}~\bibnamefont{Atat\"ure}},
  \bibinfo{journal}{Phys. Rev. Lett.} \textbf{\bibinfo{volume}{107}},
  \bibinfo{pages}{166802} (\bibinfo{year}{2011}).

\bibitem[{\citenamefont{Houel et~al.}(2012)\citenamefont{Houel, Kuhlmann,
  Greuter, Xue, Poggio, Gerardot, Dalgarno, Badolato, Petroff, Ludwig
  et~al.}}]{Houel2012}
\bibinfo{author}{\bibfnamefont{J.}~\bibnamefont{Houel}},
  \bibinfo{author}{\bibfnamefont{A.~V.} \bibnamefont{Kuhlmann}},
  \bibinfo{author}{\bibfnamefont{L.}~\bibnamefont{Greuter}},
  \bibinfo{author}{\bibfnamefont{F.}~\bibnamefont{Xue}},
  \bibinfo{author}{\bibfnamefont{M.}~\bibnamefont{Poggio}},
  \bibinfo{author}{\bibfnamefont{B.~D.} \bibnamefont{Gerardot}},
  \bibinfo{author}{\bibfnamefont{P.~A.} \bibnamefont{Dalgarno}},
  \bibinfo{author}{\bibfnamefont{A.}~\bibnamefont{Badolato}},
  \bibinfo{author}{\bibfnamefont{P.~M.} \bibnamefont{Petroff}},
  \bibinfo{author}{\bibfnamefont{A.}~\bibnamefont{Ludwig}},
  \bibnamefont{et~al.}, \bibinfo{journal}{Phys. Rev. Lett.}
  \textbf{\bibinfo{volume}{108}}, \bibinfo{pages}{107401}
  (\bibinfo{year}{2012}).

\bibitem[{\citenamefont{Hauck et~al.}(2014)\citenamefont{Hauck, Seilmeier,
  Beavan, Badolato, Petroff, and H\"ogele}}]{Hauck2014}
\bibinfo{author}{\bibfnamefont{M.}~\bibnamefont{Hauck}},
  \bibinfo{author}{\bibfnamefont{F.}~\bibnamefont{Seilmeier}},
  \bibinfo{author}{\bibfnamefont{S.~E.} \bibnamefont{Beavan}},
  \bibinfo{author}{\bibfnamefont{A.}~\bibnamefont{Badolato}},
  \bibinfo{author}{\bibfnamefont{P.~M.} \bibnamefont{Petroff}},
  \bibnamefont{and} \bibinfo{author}{\bibfnamefont{A.}~\bibnamefont{H\"ogele}},
  \bibinfo{journal}{Phys. Rev. B} \textbf{\bibinfo{volume}{90}},
  \bibinfo{pages}{235306} (\bibinfo{year}{2014}).

\bibitem[{\citenamefont{Mouri et~al.}(2013)\citenamefont{Mouri, Miyauchi,
  Iwamura, and Matsuda}}]{Mouri2013}
\bibinfo{author}{\bibfnamefont{S.}~\bibnamefont{Mouri}},
  \bibinfo{author}{\bibfnamefont{Y.}~\bibnamefont{Miyauchi}},
  \bibinfo{author}{\bibfnamefont{M.}~\bibnamefont{Iwamura}}, \bibnamefont{and}
  \bibinfo{author}{\bibfnamefont{K.}~\bibnamefont{Matsuda}},
  \bibinfo{journal}{Phys. Rev. B} \textbf{\bibinfo{volume}{87}},
  \bibinfo{pages}{045408} (\bibinfo{year}{2013}).

\bibitem[{SI()}]{SI}
\bibinfo{howpublished}{See supplementary online material for details.}

\bibitem[{\citenamefont{Bachilo et~al.}(2003)\citenamefont{Bachilo, Balzano,
  Herrera, Pompeo, Resasco, and Weisman}}]{Bachilo2003}
\bibinfo{author}{\bibfnamefont{S.~M.} \bibnamefont{Bachilo}},
  \bibinfo{author}{\bibfnamefont{L.}~\bibnamefont{Balzano}},
  \bibinfo{author}{\bibfnamefont{J.~E.} \bibnamefont{Herrera}},
  \bibinfo{author}{\bibfnamefont{F.}~\bibnamefont{Pompeo}},
  \bibinfo{author}{\bibfnamefont{D.~E.} \bibnamefont{Resasco}},
  \bibnamefont{and} \bibinfo{author}{\bibfnamefont{R.~B.}
  \bibnamefont{Weisman}}, \bibinfo{journal}{Journal of the American Chemical
  Society} \textbf{\bibinfo{volume}{125}}, \bibinfo{pages}{11186}
  (\bibinfo{year}{2003}).

\bibitem[{\citenamefont{Galland et~al.}(2008)\citenamefont{Galland, H\"ogele,
  T\"ureci, and Imamoglu}}]{Galland2008}
\bibinfo{author}{\bibfnamefont{C.}~\bibnamefont{Galland}},
  \bibinfo{author}{\bibfnamefont{A.}~\bibnamefont{H\"ogele}},
  \bibinfo{author}{\bibfnamefont{H.~E.} \bibnamefont{T\"ureci}},
  \bibnamefont{and} \bibinfo{author}{\bibfnamefont{A.}~\bibnamefont{Imamoglu}},
  \bibinfo{journal}{Phys. Rev. Lett.} \textbf{\bibinfo{volume}{101}},
  \bibinfo{pages}{067402} (\bibinfo{year}{2008}).

\bibitem[{\citenamefont{Vialla et~al.}(2014)\citenamefont{Vialla, Chassagneux,
  Ferreira, Roquelet, Diederichs, Cassabois, Roussignol, Lauret, and
  Voisin}}]{Vialla2014}
\bibinfo{author}{\bibfnamefont{F.}~\bibnamefont{Vialla}},
  \bibinfo{author}{\bibfnamefont{Y.}~\bibnamefont{Chassagneux}},
  \bibinfo{author}{\bibfnamefont{R.}~\bibnamefont{Ferreira}},
  \bibinfo{author}{\bibfnamefont{C.}~\bibnamefont{Roquelet}},
  \bibinfo{author}{\bibfnamefont{C.}~\bibnamefont{Diederichs}},
  \bibinfo{author}{\bibfnamefont{G.}~\bibnamefont{Cassabois}},
  \bibinfo{author}{\bibfnamefont{P.}~\bibnamefont{Roussignol}},
  \bibinfo{author}{\bibfnamefont{J.~S.} \bibnamefont{Lauret}},
  \bibnamefont{and} \bibinfo{author}{\bibfnamefont{C.}~\bibnamefont{Voisin}},
  \bibinfo{journal}{Phys. Rev. Lett.} \textbf{\bibinfo{volume}{113}},
  \bibinfo{pages}{057402} (\bibinfo{year}{2014}).

\bibitem[{\citenamefont{Benedict et~al.}(1995)\citenamefont{Benedict, Louie,
  and Cohen}}]{Benedict1995}
\bibinfo{author}{\bibfnamefont{L.~X.} \bibnamefont{Benedict}},
  \bibinfo{author}{\bibfnamefont{S.~G.} \bibnamefont{Louie}}, \bibnamefont{and}
  \bibinfo{author}{\bibfnamefont{M.~L.} \bibnamefont{Cohen}},
  \bibinfo{journal}{Phys. Rev. B} \textbf{\bibinfo{volume}{52}},
  \bibinfo{pages}{8541} (\bibinfo{year}{1995}).

\bibitem[{\citenamefont{Li et~al.}(2003)\citenamefont{Li, Rotkin, and
  Ravaioli}}]{Li2003}
\bibinfo{author}{\bibfnamefont{Y.}~\bibnamefont{Li}},
  \bibinfo{author}{\bibfnamefont{S.~V.} \bibnamefont{Rotkin}},
  \bibnamefont{and} \bibinfo{author}{\bibfnamefont{U.}~\bibnamefont{Ravaioli}},
  \bibinfo{journal}{Nano Lett.} \textbf{\bibinfo{volume}{3}},
  \bibinfo{pages}{183} (\bibinfo{year}{2003}).

\bibitem[{\citenamefont{Novikov and Levitov}(2006)}]{Novikov2006}
\bibinfo{author}{\bibfnamefont{D.~S.} \bibnamefont{Novikov}} \bibnamefont{and}
  \bibinfo{author}{\bibfnamefont{L.~S.} \bibnamefont{Levitov}},
  \bibinfo{journal}{Phys. Rev. Lett.} \textbf{\bibinfo{volume}{96}},
  \bibinfo{pages}{036402} (\bibinfo{year}{2006}).

\bibitem[{\citenamefont{Guo et~al.}(2004)\citenamefont{Guo, Chu, Wang, and
  Duan}}]{Guo2004}
\bibinfo{author}{\bibfnamefont{G.~Y.} \bibnamefont{Guo}},
  \bibinfo{author}{\bibfnamefont{K.~C.} \bibnamefont{Chu}},
  \bibinfo{author}{\bibfnamefont{D.~S.} \bibnamefont{Wang}}, \bibnamefont{and}
  \bibinfo{author}{\bibfnamefont{C.~G.} \bibnamefont{Duan}},
  \bibinfo{journal}{Phys. Rev. B} \textbf{\bibinfo{volume}{69}},
  \bibinfo{pages}{205416} (\bibinfo{year}{2004}).

\bibitem[{\citenamefont{Brothers et~al.}(2005)\citenamefont{Brothers, Kudin,
  Scuseria, and Bauschlicher}}]{Brothers2005}
\bibinfo{author}{\bibfnamefont{E.~N.} \bibnamefont{Brothers}},
  \bibinfo{author}{\bibfnamefont{K.~N.} \bibnamefont{Kudin}},
  \bibinfo{author}{\bibfnamefont{G.~E.} \bibnamefont{Scuseria}},
  \bibnamefont{and} \bibinfo{author}{\bibfnamefont{C.~W.}
  \bibnamefont{Bauschlicher}}, \bibinfo{journal}{Phys. Rev. B}
  \textbf{\bibinfo{volume}{72}}, \bibinfo{pages}{033402}
  (\bibinfo{year}{2005}).

\bibitem[{\citenamefont{Kozinsky and Marzari}(2006)}]{Kozinsky2006}
\bibinfo{author}{\bibfnamefont{B.}~\bibnamefont{Kozinsky}} \bibnamefont{and}
  \bibinfo{author}{\bibfnamefont{N.}~\bibnamefont{Marzari}},
  \bibinfo{journal}{Phys. Rev. Lett.} \textbf{\bibinfo{volume}{96}},
  \bibinfo{pages}{166801} (\bibinfo{year}{2006}).

\bibitem[{\citenamefont{Santos et~al.}(2011)\citenamefont{Santos, Yuma,
  Berciaud, Shaver, Gallart, Gilliot, Cognet, and Lounis}}]{Santos2011}
\bibinfo{author}{\bibfnamefont{S.~M.} \bibnamefont{Santos}},
  \bibinfo{author}{\bibfnamefont{B.}~\bibnamefont{Yuma}},
  \bibinfo{author}{\bibfnamefont{S.}~\bibnamefont{Berciaud}},
  \bibinfo{author}{\bibfnamefont{J.}~\bibnamefont{Shaver}},
  \bibinfo{author}{\bibfnamefont{M.}~\bibnamefont{Gallart}},
  \bibinfo{author}{\bibfnamefont{P.}~\bibnamefont{Gilliot}},
  \bibinfo{author}{\bibfnamefont{L.}~\bibnamefont{Cognet}}, \bibnamefont{and}
  \bibinfo{author}{\bibfnamefont{B.}~\bibnamefont{Lounis}},
  \bibinfo{journal}{Phys. Rev. Lett.} \textbf{\bibinfo{volume}{107}},
  \bibinfo{pages}{187401} (\bibinfo{year}{2011}).

\bibitem[{\citenamefont{Yoshida et~al.}(2016)\citenamefont{Yoshida, Popert, and
  Kato}}]{Yoshida2016}
\bibinfo{author}{\bibfnamefont{M.}~\bibnamefont{Yoshida}},
  \bibinfo{author}{\bibfnamefont{A.}~\bibnamefont{Popert}}, \bibnamefont{and}
  \bibinfo{author}{\bibfnamefont{Y.~K.} \bibnamefont{Kato}},
  \bibinfo{journal}{Phys. Rev. B} \textbf{\bibinfo{volume}{93}},
  \bibinfo{pages}{041402} (\bibinfo{year}{2016}).

\bibitem[{\citenamefont{Ronnow et~al.}(2009)\citenamefont{Ronnow, Pedersen, and
  Cornean}}]{Ronnow2009}
\bibinfo{author}{\bibfnamefont{T.~F.} \bibnamefont{Ronnow}},
  \bibinfo{author}{\bibfnamefont{T.~G.} \bibnamefont{Pedersen}},
  \bibnamefont{and} \bibinfo{author}{\bibfnamefont{H.~D.}
  \bibnamefont{Cornean}}, \bibinfo{journal}{Phys. Lett. A}
  \textbf{\bibinfo{volume}{373}}, \bibinfo{pages}{1478} (\bibinfo{year}{2009}).

\bibitem[{\citenamefont{Warburton et~al.}(2000)\citenamefont{Warburton,
  Sch{\"a}flein, Haft, Bickel, Lorke, Karrai, Garcia, Schoenfeld, and
  Petroff}}]{Warburton2000}
\bibinfo{author}{\bibfnamefont{R.~J.} \bibnamefont{Warburton}},
  \bibinfo{author}{\bibfnamefont{C.}~\bibnamefont{Sch{\"a}flein}},
  \bibinfo{author}{\bibfnamefont{D.}~\bibnamefont{Haft}},
  \bibinfo{author}{\bibfnamefont{F.}~\bibnamefont{Bickel}},
  \bibinfo{author}{\bibfnamefont{A.}~\bibnamefont{Lorke}},
  \bibinfo{author}{\bibfnamefont{K.}~\bibnamefont{Karrai}},
  \bibinfo{author}{\bibfnamefont{J.~M.} \bibnamefont{Garcia}},
  \bibinfo{author}{\bibfnamefont{W.}~\bibnamefont{Schoenfeld}},
  \bibnamefont{and} \bibinfo{author}{\bibfnamefont{P.~M.}
  \bibnamefont{Petroff}}, \bibinfo{journal}{Nature}
  \textbf{\bibinfo{volume}{405}}, \bibinfo{pages}{926} (\bibinfo{year}{2000}).

\bibitem[{\citenamefont{Gruber et~al.}(1997)\citenamefont{Gruber,
  Dr{\"a}benstedt, Tietz, Fleury, Wrachtrup, and Borczyskowski}}]{Gruber1997}
\bibinfo{author}{\bibfnamefont{A.}~\bibnamefont{Gruber}},
  \bibinfo{author}{\bibfnamefont{A.}~\bibnamefont{Dr{\"a}benstedt}},
  \bibinfo{author}{\bibfnamefont{C.}~\bibnamefont{Tietz}},
  \bibinfo{author}{\bibfnamefont{L.}~\bibnamefont{Fleury}},
  \bibinfo{author}{\bibfnamefont{J.}~\bibnamefont{Wrachtrup}},
  \bibnamefont{and} \bibinfo{author}{\bibfnamefont{C.~v.}
  \bibnamefont{Borczyskowski}}, \bibinfo{journal}{Science}
  \textbf{\bibinfo{volume}{276}}, \bibinfo{pages}{2012} (\bibinfo{year}{1997}).

\bibitem[{\citenamefont{Balasubramanian
  et~al.}(2009)\citenamefont{Balasubramanian, Neumann, Twitchen, Markham,
  Kolesov, Mizuochi, Isoya, Achard, Beck, Tissler
  et~al.}}]{Balasubramanian2009iso}
\bibinfo{author}{\bibfnamefont{G.}~\bibnamefont{Balasubramanian}},
  \bibinfo{author}{\bibfnamefont{P.}~\bibnamefont{Neumann}},
  \bibinfo{author}{\bibfnamefont{D.}~\bibnamefont{Twitchen}},
  \bibinfo{author}{\bibfnamefont{M.}~\bibnamefont{Markham}},
  \bibinfo{author}{\bibfnamefont{R.}~\bibnamefont{Kolesov}},
  \bibinfo{author}{\bibfnamefont{N.}~\bibnamefont{Mizuochi}},
  \bibinfo{author}{\bibfnamefont{J.}~\bibnamefont{Isoya}},
  \bibinfo{author}{\bibfnamefont{J.}~\bibnamefont{Achard}},
  \bibinfo{author}{\bibfnamefont{J.}~\bibnamefont{Beck}},
  \bibinfo{author}{\bibfnamefont{J.}~\bibnamefont{Tissler}},
  \bibnamefont{et~al.}, \bibinfo{journal}{Nat. Mater.}
  \textbf{\bibinfo{volume}{8}}, \bibinfo{pages}{383} (\bibinfo{year}{2009}).

\bibitem[{\citenamefont{Mamin et~al.}(2013)\citenamefont{Mamin, Kim, Sherwood,
  Rettner, Ohno, Awschalom, and Rugar}}]{Mamin2013}
\bibinfo{author}{\bibfnamefont{H.~J.} \bibnamefont{Mamin}},
  \bibinfo{author}{\bibfnamefont{M.}~\bibnamefont{Kim}},
  \bibinfo{author}{\bibfnamefont{M.~H.} \bibnamefont{Sherwood}},
  \bibinfo{author}{\bibfnamefont{C.~T.} \bibnamefont{Rettner}},
  \bibinfo{author}{\bibfnamefont{K.}~\bibnamefont{Ohno}},
  \bibinfo{author}{\bibfnamefont{D.~D.} \bibnamefont{Awschalom}},
  \bibnamefont{and} \bibinfo{author}{\bibfnamefont{D.}~\bibnamefont{Rugar}},
  \bibinfo{journal}{Science} \textbf{\bibinfo{volume}{339}},
  \bibinfo{pages}{557} (\bibinfo{year}{2013}).

\bibitem[{\citenamefont{Staudacher et~al.}(2013)\citenamefont{Staudacher, Shi,
  Pezzagna, Meijer, Du, Meriles, Reinhard, and Wrachtrup}}]{Staudacher2013}
\bibinfo{author}{\bibfnamefont{T.}~\bibnamefont{Staudacher}},
  \bibinfo{author}{\bibfnamefont{F.}~\bibnamefont{Shi}},
  \bibinfo{author}{\bibfnamefont{S.}~\bibnamefont{Pezzagna}},
  \bibinfo{author}{\bibfnamefont{J.}~\bibnamefont{Meijer}},
  \bibinfo{author}{\bibfnamefont{J.}~\bibnamefont{Du}},
  \bibinfo{author}{\bibfnamefont{C.~A.} \bibnamefont{Meriles}},
  \bibinfo{author}{\bibfnamefont{F.}~\bibnamefont{Reinhard}}, \bibnamefont{and}
  \bibinfo{author}{\bibfnamefont{J.}~\bibnamefont{Wrachtrup}},
  \bibinfo{journal}{Science} \textbf{\bibinfo{volume}{339}},
  \bibinfo{pages}{561} (\bibinfo{year}{2013}).

\bibitem[{\citenamefont{Laraoui et~al.}(2012)\citenamefont{Laraoui, Hodges, and
  Meriles}}]{Laraoui2012}
\bibinfo{author}{\bibfnamefont{A.}~\bibnamefont{Laraoui}},
  \bibinfo{author}{\bibfnamefont{J.~S.} \bibnamefont{Hodges}},
  \bibnamefont{and} \bibinfo{author}{\bibfnamefont{C.~A.}
  \bibnamefont{Meriles}}, \bibinfo{journal}{Nano Lett.}
  \textbf{\bibinfo{volume}{12}}, \bibinfo{pages}{3477} (\bibinfo{year}{2012}).

\bibitem[{\citenamefont{Wang et~al.}(2012)\citenamefont{Wang, Kalantar-Zadeh,
  Kis, Coleman, and Strano}}]{Wang2012}
\bibinfo{author}{\bibfnamefont{Q.~H.} \bibnamefont{Wang}},
  \bibinfo{author}{\bibfnamefont{K.}~\bibnamefont{Kalantar-Zadeh}},
  \bibinfo{author}{\bibfnamefont{A.}~\bibnamefont{Kis}},
  \bibinfo{author}{\bibfnamefont{J.~N.} \bibnamefont{Coleman}},
  \bibnamefont{and} \bibinfo{author}{\bibfnamefont{M.~S.}
  \bibnamefont{Strano}}, \bibinfo{journal}{Nat. Nanotechnol.}
  \textbf{\bibinfo{volume}{7}}, \bibinfo{pages}{699} (\bibinfo{year}{2012}).

\end{thebibliography}

\clearpage

\begin{thebibliography}{9}
\expandafter\ifx\csname
natexlab\endcsname\relax\def\natexlab#1{#1}\fi
\expandafter\ifx\csname bibnamefont\endcsname\relax
  \def\bibnamefont#1{#1}\fi
\expandafter\ifx\csname bibfnamefont\endcsname\relax
  \def\bibfnamefont#1{#1}\fi
\expandafter\ifx\csname citenamefont\endcsname\relax
  \def\citenamefont#1{#1}\fi
\expandafter\ifx\csname url\endcsname\relax
  \def\url#1{\texttt{#1}}\fi
\expandafter\ifx\csname
urlprefix\endcsname\relax\def\urlprefix{URL }\fi
\providecommand{\bibinfo}[2]{#2}
\providecommand{\eprint}[2][]{\url{#2}}

\bibitem[{\citenamefont{Sze}(1981)}]{Sze}
\bibinfo{author}{\bibfnamefont{S.~M.} \bibnamefont{Sze}},
  \emph{\bibinfo{title}{Physics of Semiconductor Devices}}
  (\bibinfo{publisher}{John Wiley \& Sons}, \bibinfo{year}{1981}),
  \bibinfo{edition}{2nd} ed.

\bibitem[{\citenamefont{Grove et~al.}(1965)\citenamefont{Grove, Deal, Snow, and
  Sah}}]{Grove}
\bibinfo{author}{\bibfnamefont{A.}~\bibnamefont{Grove}},
  \bibinfo{author}{\bibfnamefont{B.}~\bibnamefont{Deal}},
  \bibinfo{author}{\bibfnamefont{E.}~\bibnamefont{Snow}}, \bibnamefont{and}
  \bibinfo{author}{\bibfnamefont{C.}~\bibnamefont{Sah}},
  \bibinfo{journal}{Solid-State Electronics} \textbf{\bibinfo{volume}{8}},
  \bibinfo{pages}{145} (\bibinfo{year}{1965}).

\bibitem[{\citenamefont{Goetzberger. and Irvin}(1968)}]{Goetzberger}
\bibinfo{author}{\bibfnamefont{A.}~\bibnamefont{Goetzberger.}}
  \bibnamefont{and} \bibinfo{author}{\bibfnamefont{J.~C.} \bibnamefont{Irvin}},
  \bibinfo{journal}{IEEE Trans. Electron Dev.} \textbf{\bibinfo{volume}{15}},
  \bibinfo{pages}{1009} (\bibinfo{year}{1968}).

\bibitem[{\citenamefont{Frisch et~al.}(2015)\citenamefont{Frisch, Trucks,
  Schlegel, Scuseria, Robb, Cheeseman, Scalmani, Barone, Mennucci, Petersson
  et~al.}}]{Frisch2009}
\bibinfo{author}{\bibfnamefont{M.}~\bibnamefont{Frisch}},
  \bibinfo{author}{\bibfnamefont{G.}~\bibnamefont{Trucks}},
  \bibinfo{author}{\bibfnamefont{H.}~\bibnamefont{Schlegel}},
  \bibinfo{author}{\bibfnamefont{G.}~\bibnamefont{Scuseria}},
  \bibinfo{author}{\bibfnamefont{M.}~\bibnamefont{Robb}},
  \bibinfo{author}{\bibfnamefont{J.}~\bibnamefont{Cheeseman}},
  \bibinfo{author}{\bibfnamefont{G.}~\bibnamefont{Scalmani}},
  \bibinfo{author}{\bibfnamefont{V.}~\bibnamefont{Barone}},
  \bibinfo{author}{\bibfnamefont{B.}~\bibnamefont{Mennucci}},
  \bibinfo{author}{\bibfnamefont{G.}~\bibnamefont{Petersson}},
  \bibnamefont{et~al.}, \bibinfo{journal}{Gaussian 09, Revision A. 02;
  Gaussian, Inc: Wallingford, CT, 2009}  (\bibinfo{year}{2015}).

\bibitem[{\citenamefont{Becke}(1988)}]{Becke1988}
\bibinfo{author}{\bibfnamefont{A.~D.} \bibnamefont{Becke}},
  \bibinfo{journal}{Phys. Rev. A} \textbf{\bibinfo{volume}{38}},
  \bibinfo{pages}{3098} (\bibinfo{year}{1988}).

\bibitem[{\citenamefont{Cossi et~al.}(2003)\citenamefont{Cossi, Rega, Scalmani,
  and Barone}}]{Cossi2003}
\bibinfo{author}{\bibfnamefont{M.}~\bibnamefont{Cossi}},
  \bibinfo{author}{\bibfnamefont{N.}~\bibnamefont{Rega}},
  \bibinfo{author}{\bibfnamefont{G.}~\bibnamefont{Scalmani}}, \bibnamefont{and}
  \bibinfo{author}{\bibfnamefont{V.}~\bibnamefont{Barone}},
  \bibinfo{journal}{J. Comput. Chem.} \textbf{\bibinfo{volume}{24}},
  \bibinfo{pages}{669} (\bibinfo{year}{2003}).

\bibitem[{\citenamefont{Barone and Cossi}(1998)}]{Barone1998}
\bibinfo{author}{\bibfnamefont{V.}~\bibnamefont{Barone}} \bibnamefont{and}
  \bibinfo{author}{\bibfnamefont{M.}~\bibnamefont{Cossi}}, \bibinfo{journal}{J.
  Phys. Chem. A} \textbf{\bibinfo{volume}{102}}, \bibinfo{pages}{1995}
  (\bibinfo{year}{1998}).

\bibitem[{\citenamefont{Mukamel et~al.}(1997)\citenamefont{Mukamel, Tretiak,
  Wagersreiter, and Chernyak}}]{Mukamel1997}
\bibinfo{author}{\bibfnamefont{S.}~\bibnamefont{Mukamel}},
  \bibinfo{author}{\bibfnamefont{S.}~\bibnamefont{Tretiak}},
  \bibinfo{author}{\bibfnamefont{T.}~\bibnamefont{Wagersreiter}},
  \bibnamefont{and} \bibinfo{author}{\bibfnamefont{V.}~\bibnamefont{Chernyak}},
  \bibinfo{journal}{Science} \textbf{\bibinfo{volume}{277}},
  \bibinfo{pages}{781} (\bibinfo{year}{1997}).

\bibitem[{\citenamefont{Tretiak and Mukamel}(2002)}]{Tretiak2002}
\bibinfo{author}{\bibfnamefont{S.}~\bibnamefont{Tretiak}} \bibnamefont{and}
  \bibinfo{author}{\bibfnamefont{S.}~\bibnamefont{Mukamel}},
  \bibinfo{journal}{Chem. Rev.} \textbf{\bibinfo{volume}{102}},
  \bibinfo{pages}{3171} (\bibinfo{year}{2002}).

\end{thebibliography}
\end{document}